# Current-driven domain wall dynamics in ferromagnetic layers synthetically exchange-coupled by a spacer: A micromagnetic study

Oscar Alejos[2], Victor Raposo[1], Luis Sanchez-Tejerina[2], Riccardo Tomasello[3], Giovanni Finocchio[4], and Eduardo Martinez[1,*]

[1]Dpto. Fisica Aplicada, University of Salamanca, 37008 Salamanca, Spain.
[2]Dpto. Electricidad y Electronica, University of Valladolid, 47011 Valladolid, Spain.
[3]Dept. of Engineering, University of Perugia, 06123 Perugia, Italy
[4]Dept. of Mathematical and Computer Sciences, Physical Sciences and Earth Sciences, University of Messina, 98122 Messina, Italy
* corresponding author: edumartinez@usal.es

## Abstract

The current-driven domain wall motion along two exchange-coupled ferromagnetic layers with perpendicular anisotropy is studied by means of micromagnetic simulations and compared to the conventional case of a single ferromagnetic layer. Our results, where only the lower ferromagnetic layer is subjected to the interfacial Dzyaloshinskii-Moriya interaction and to the spin Hall effect, indicate that the domain walls can be synchronously driven in the presence of a strong interlayer exchange coupling, and that the velocity is significantly enhanced due to the antiferromagnetic exchange coupling as compared with the single-layer case. On the contrary, when the coupling is of ferromagnetic nature, the velocity is reduced. We provide a full micromagnetic characterization of the current-driven motion in these multilayers, both in the absence and in the presence of longitudinal fields, and the results are explained based on a one-dimensional model. The interfacial Dzyaloshinskii-Moriya interaction, only necessary in this lower layer, gives the required chirality to the magnetization textures, while the interlayer exchange coupling favors the synchronous movement of the coupled walls by a dragging mechanism, without significant tilting of the domain wall plane. Finally, the domain wall dynamics along curved strips is also evaluated. These results indicate that the antiferromagnetic coupling between the ferromagnetic layers mitigates the tilting of the walls, which suggest these systems to achieve efficient and highly-packed displacement of trains of walls for spintronics devices. A study, taking into account defects and thermal fluctuations, allows to analyze the validity range of these claims.

## 1. INTRODUCTION

Understanding and controlling the dynamics of domain walls (DWs) along ultrathin magnetic heterostructures consisting of a ferromagnetic (FM) strip sandwiched between a heavy metal (HM) and an Oxide is nowadays the focus of intense research[1,2,3,4,5,6,7,8]. These HM/FM/Oxide multilayers exhibit high perpendicular magnetocrystalline anisotropy (PMA), and the broken inversion symmetry at the interfaces promotes chiral Néel walls by the Dzyaloshinskii-Moriya interaction (DMI)[4,5,6,7,8,9,10,11]. These DWs can be efficiently driven by current pulses as due to the spin Hall effect (SHE) in the HM[4,5,12,13].

Recent experimental works[14,15] have shown that the DW dynamics can be even optimized in synthetic antiferromagnetic heterostructures (SAF), where antiferromagnetic coupling appears between two ferromagnetic layers isolated by means of a non-magnetic spacer. The whole heterostructure can be represented by HM/LFM/Spacer/UFM, where LFM and UFM stand for the lower and the upper FM layers respectively. In these two-FM layers heterostructures, the DWs can be displaced even more efficiently and at much higher speeds if compared with the single-FM-layer stack (HM/FM/Oxide). This is due to a stabilization of the Néel DW configuration, and the exchange coupling torque that is directly proportional to the strength of the antiferromagnetic exchange coupling between the two FM layers. Moreover, because of the exchange coupling torque, the dependence of the DW velocity on the magnetic field applied along the nanowire is different from that of the single-FM-layer heterostructure. These experimental results[14] were explained within the framework of a one-dimensional model (1DM), which deals with the dynamics of the coupled DWs in the LFM and in the UFM layers by considering them as rigid objects. However, a more realistic analysis, taking into account the full three-dimensional dependence of the magnetization in the two FM layers, is still missing and needed to test the validity of the 1DM.

Accordingly, the current-driven DW (CDDW) dynamics in HM/LFM/Spacer/UFM multilayers is here investigated by means of full micromagnetic ($\mu M$) simulations, and compared with the behavior of a single-FM-layer stack (HM/FM/Oxide). The two considered multilayer systems are sketched in Fig. 1(a) and (b), and the description of the geometry and dimensions are given in its caption. To elucidate the relevant aspects of this CDDW dynamics, simulations consider perfect strips as a first approach, although additional simulations, which mimic realistic conditions by including disorder and thermal effects, have been also carried out. In order to collect detailed information about the acting mechanisms associated to the coupling between the FM layers through the spacer, which is determined by a certain interlayer exchange parameter $J^{ex}$[16,17], the magnetization state of both FM layers is simultaneously evaluated. Finally, both ferromagnetic (FM) and antiferromagnetic (AF) coupling cases are considered, the former given by a positive $J^{ex}$, and the latter by a negative $J^{ex}$.

(a) Two FM layers heteroestructure (HM/LFM/S/UFM)

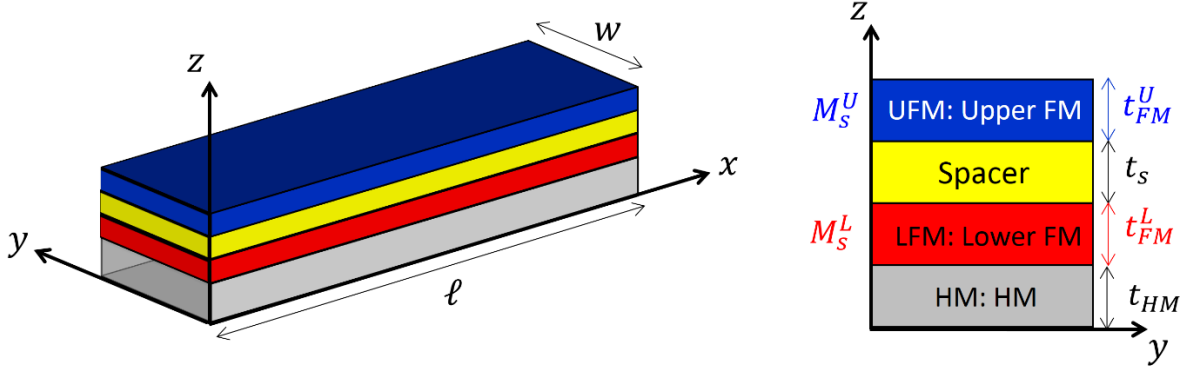

(b) Single FM layer stack (HM/FM)

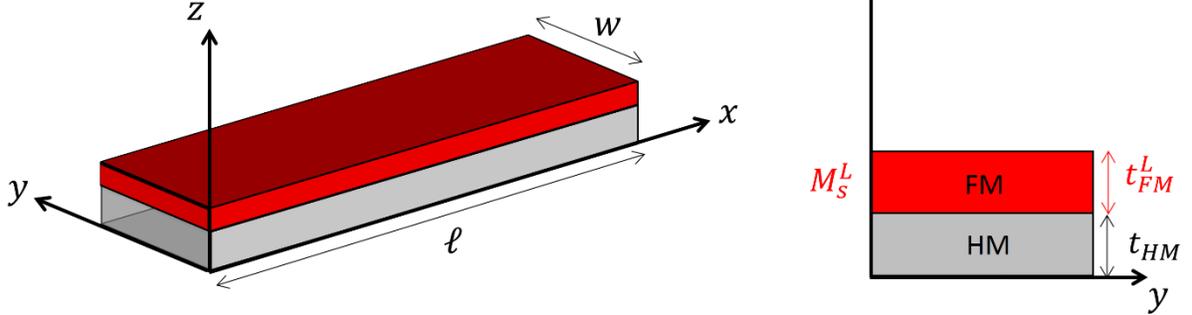

**Figure 1.** (a) Schematic representation of the multilayer structure with two FM layers. The relevant thicknesses for this study are marked on the figure, which are fixed to $t_{FM}^L = t_s = t_{FM}^U = 0.8$ nm, except otherwise indicated. The saturation magnetizations are also fixed by default to $M_S^L = 600$ kA/m and $M_S^U = 600$ kA/m. The width is $w = 100$ nm. The anisotropy constant, the intralayer exchange constant and the Gilbert damping are respectively $K_1 = 0.6$ MJ/m³, $A = 20$ pJ/m and $\alpha = 0.1$ for both FM layers. The interfacial DMI in the lower FM is $D^L = 1.25$ mJ/m², and no DMI is considered for the UFM ($D^U = 0$). Along this work, the single-FM-layer (b) is considered to have identical characteristics to those of the lower FM layer. Parameters here used can be found in the literature[4,7,14].

The manuscript is structured as follows. Section 2 describes the details of the micromagnetic model ($\mu M$) and the one-dimensional model (1DM). The current-driven DW dynamics along perfect samples, both in the absence and in the presence of in-plane longitudinal fields, is presented in Section 3. Micromagnetic results for realistic samples are presented in Section 4 for different multilayers where the thickness of the FM layers and the spacer ($t_{FM}^L$, $t_{FM}^U$ and $t_S$) and the saturation magnetization of the layers ($M_S^L$ and $M_S^U$) are varied. The current-driven DW motion along multilayers with curved parts is studied in Section 5, and the main conclusions are discussed in Section 6.

## 2. MODELS AND NUMERICAL DETAILS.

### 2.1. Micromagnetic model ($\mu M$)

Full micromagnetic ($\mu M$) simulations have been performed by solving the Landau-Lifshitz-Gilbert equation augmented with the spin transfer torque ($\vec{\tau}_{ST}$, STT) and Slonczewski-like spin-orbit torque ($\vec{\tau}_{SO}$, SOT) due to the spin Hall torque[18,19]:

$$\frac{d\vec{m}}{dt} = -\gamma_0 \vec{m} \times (\vec{H}_{eff} + \vec{H}_{th}) + \alpha \vec{m} \times \frac{d\vec{m}}{dt} + \vec{\tau}_{ST} + \vec{\tau}_{SO} \qquad (2)$$

where $\gamma_0$ and $\alpha$ are the gyromagnetic ratio and the Gilbert damping constant respectively. $\vec{m}(\vec{r},t) \equiv \vec{m}^i(\vec{r},t) = \vec{M}^i(\vec{r},t)/M_S^i$ is the normalized local magnetization to its saturation value ($M_S^i$), defined differently for each FM layer: $M_S^i$ where $i: L, U$ for the LFM and the UFM layers respectively. $\vec{H}_{eff}$ is the deterministic effective field, which includes not only the intralayer exchange and the uniaxial anisotropy, but also the interlayer exchange[17] and the magnetostatic interactions adequately weighed to account for the different saturation magnetizations. The interlayer exchange contribution ($\vec{H}_{ex}^{inter}$) to the effective field $\vec{H}_{eff}$, acting on each FM layer, is computed from the corresponding energy density ($\omega_{ex}^{inter} = -\frac{J^{ex}}{t_S} \vec{m}^L \cdot \vec{m}^U$, where $J^{ex}$ is the interlayer exchange coupling parameter, $t_S$ is the thickness of the spacer between the LFM and the UFM layers, and $\vec{m}^L$ and $\vec{m}^U$ represent the normalized magnetization in the Lower and in the Upper layers respectively) as

$$\vec{H}_{ex,i}^{inter} = -\frac{1}{\mu_0 M_S^i} \frac{\delta \omega_{ex}^{inter}}{\delta \vec{m}^i} = \frac{J^{ex}}{\mu_0 M_S^j t_S} \vec{m}^j \qquad (3)$$

where $i, j: L, U$. Ferromagnetic (FM) and antiferromagnetic (AF) coupling cases are evaluated by a positive $J^{ex}$, and by a negative $J^{ex}$ respectively.

The effective field in the LFM layer requires an additional term representing the interfacial DMI at the HM/LFM interface. The rest of numerical details of other contributions to the effective field can be found elsewhere[18]. $\vec{H}_{th}$ is the thermal field, included as a Gaussian-distributed random field[20,21]. $\vec{\tau}_{SO}$ represents the spin-orbit torque (SOT), which in the present work is solely acting on the LFM layer (the only one contacting the HM). This torque is given by the Slonczewski-like term $\vec{\tau}_{SO} = -\gamma_0 \vec{m} \times \vec{H}_{SL}$, where $\vec{H}_{SL} = H_{SL}^0 \vec{m} \times \vec{\sigma}$ is the Slonczewski-like effective field. Here, $\vec{\sigma} = \vec{u}_z \times \vec{u}_J$ is the unit vector along the direction of the polarization of the spin current generated by the spin Hall effect (SHE) in the HM, being orthogonal to both the direction of the electric current $\vec{u}_J$ and the vector $\vec{u}_z$ standing for the normal to the HM/LFM interface. Finally, $H_{SL}^0 = \hbar \theta_{SH} J_{HM} / (2\mu_0 |e| M_s t_{FM})$ determines the strength of the SHE[4], where $\hbar$ is the Planck constant, $e$ is the electron charge, $\mu_0$ is the vacuum permeability, $\theta_{SH}$ is the spin Hall angle, and $J$ is the magnitude of the current density $\vec{J}_{HM}(t) = J_{HM}(t) \vec{u}_J$. For straight samples $\vec{u}_J = \vec{u}_x$, whereas for curved strips the direction and the local amplitude of current was previously computed by finite element method solvers[22]. On the other hand, Eq. (2) includes the spin transfer torques (STTs, $\vec{\tau}_{ST}$) due to the electrical current flowing across the FM layers ($\vec{J}_{HM}^i(t) = J_{FM}^i(t) \vec{u}_J$, with $i = \{U, L\}$ ). This STTs[21] includes both adiabatic and non-adiabatic contributions: $\vec{\tau}_{ST} = b_{ST}^i (\vec{u}_J \cdot \nabla) \vec{m} - \xi_i b_{ST}^i \vec{m} \times (\vec{u}_J \cdot \nabla) \vec{m}$, where $b_{ST}^i = \frac{\mu_B P}{|e| M_s^i} J_{FM}^i$ is the STT coefficient, with $P$ the polarization factor and $J_{FM}^i$ the density current flowing directly throw the FM layer $i = \{U, L\}$. $\xi_i$ is the non-adiabatic coefficient[21].

Typical values of the parameters above have been considered in our simulations. Except where the contrary is indicated, $M_s$ values for the UFM and the LFM layers have been chosen respectively as $M_s^L = 600$ kA/m and $M_s^U = 600$ kA/m. The anisotropy constant, the intralayer exchange constant and the Gilbert damping are $K_u = 0.6$ MJ/m³, $A = 20$ pJ/m and $\alpha = 0.1$ for both FM layers. The interfacial DMI in the lower FM is $D^L = 1.25$ mJ/m², and null in the upper FM layer ($D^U = 0$). The spin Hall angle representing the degree of polarization of the vertical spin current acting on the LFM is $\theta_{SH} = -0.12$. Different values of the interlayer exchange parameter ($J^{ex}$) have been considered with magnitudes within the range $0 \le |J^{ex}| \le 0.5$ mJ/m², but with $J^{ex}$ taking by default the values $J^{ex} = \pm 0.5$ mJ/m² for FM and AF coupling cases. For the study of single-FM layer, the following parameters were adopted: $M_S = 600$ kA/m, $K_u = 0.6$ MJ/m$^3$, $A = 20$ pJ/m, $\alpha = 0.1$, $D = 1.25$ mJ/m$^2$ and $\theta_{SH} = -0.12$, which coincide with the ones chosen for the LFM layer in the two FM layer cases. Initially, we assume that STT is negligible ($b_{ST}^i = 0$) in the evaluated samples. We will show in Sec. 3.1.B that indeed the STT plays a marginal role on the current-driven dynamics evaluated in the present study. The dynamics equation of the magnetization over the full system was solved using MuMax3[23] which was adapted to include the Ruderman–Kittel–Kasuya–Yosida interaction[17] between non-adjacent FM layers separated by the spacer. The in-plane side of the computational cells is $\Delta x = \Delta y = 3$ nm and different thicknesses $\Delta z$, depending on the thickness of the FM layers, were considered. A homemade micromagnetic solver was also used to verify the validity of the obtained results. Except the contrary is said, the presented results were obtained at zero temperature. Simulations at room temperature were performed with a fixed time step $\Delta t = 0.1$ ps. Several tests were performed with reduced cell sizes and time steps to assess the numerical validity of the presented results.

Part of the simulations were carried out by considering perfect samples, without imperfections nor defects. However, other parts were computed under realistic conditions (see Sections 4 and 5). In order to take into account the effects of disorder due to imperfections and defects in a realistic way, we assume that the easy axis anisotropy direction is distributed among a length scale defined by a '*grain size*'. The grains vary in size taking an average size of 10 nm. The direction of the uniaxial anisotropy of each grain is mainly directed along the perpendicular direction ($z$-axis) but with a small in-plane component which is randomly generated over the grains. The maximum percentage of the in-plane component of the uniaxial anisotropy unit vector is varied from 10% to 15%. The presented results correspond to an in-plane maximum deviation from the out-of-plane direction of 12%. Although other ways to account for imperfection could be adopted, we selected this one based on previous studies, which properly describe other experimental observations[19].

### 2.2. One-dimensional Model (1DM)

The one dimensional model (1DM) assumes that the DW profile can be described by the Bloch's ansatz[9], and therefore its dynamics can be described by means of the DW position ($q$) and the internal DW angle ($\Phi$). The 1DM has been developed by several authors to account for and describe the field-driven and current-driven DW dynamics in different systems[4,9,18]. Yang *et al.*[14] developed this 1DM to describe the DW dynamics in bi-layer FM systems in the presence of a strong interlayer exchange coupling between the two FM layers. These Eqs. assume that the DWs in the LFM and the UFM layers move completely coupled to each other, and therefore, $q_L = q_U = q$ represents the DW position along the longitudinal axis of the two walls. The same DW width in the two FM layers was also assumed ($\Delta$). On the other hand, the internal DW angle is different for each layer: $\Phi_i$, with the index $i = \{U, L\}$ standing for the UFM and the LFM layers. We have derived the 1DM equations, which can be written as:

$$(\alpha_L M_s^L + \alpha_U M_s^U)\frac{\dot{q}}{\Delta} + Q_L M_s^L \dot{\Phi}_L + Q_U M_s^U \dot{\Phi}_U$$

$$= \gamma_0 \frac{\pi}{2} Q_U M_s^L H_{SL}^L \cos\Phi_L + \gamma_0 \frac{\pi}{2} Q_L M_s^U H_{SL}^U \cos\Phi_U \qquad (4)$$

$$+ (\gamma_0 Q_L M_s^L + \gamma_0 Q_U M_s^U)H_z - \xi_L \frac{b_{ST}^L}{\Delta} M_s^L - \xi_U \frac{b_{ST}^U}{\Delta} M_s^U,$$

$$Q_L \frac{\dot{q}}{\Delta} - \alpha_L \dot{\Phi}_L = \gamma_0 \left[ \frac{\pi}{2} H_x \sin\Phi_L - \frac{\pi}{2} H_y \cos\Phi_L - \frac{\pi}{2} Q_L H_D^L \sin\Phi_L - H_k^L \frac{\sin 2\Phi_L}{2} \right.$$

$$\left. + \frac{2J^{ex}}{\mu_0 M_s^L t_s} \sin(\Phi_L - \Phi_U) \right] - Q_L \frac{b_{ST}^L}{\Delta}, \qquad (5)$$

$$Q_U \frac{\dot{q}}{\Delta} - \alpha_U \dot{\Phi}_U = \gamma_0 \left[ \frac{\pi}{2} H_x \sin\Phi_U - \frac{\pi}{2} H_y \cos\Phi_U - \frac{\pi}{2} Q_U H_D^U \sin\Phi_U \right.$$

$$\left. - H_k^U \frac{\sin 2\Phi_U}{2} - \frac{2J^{ex}}{\mu_0 M_s^U t_s} \sin(\Phi_L - \Phi_U) \right] - Q_U \frac{b_{ST}^U}{\Delta}, \qquad (6)$$

where the top dot notation represents the time derivative ($\dot{q} \equiv \frac{dq}{dt}$), and $Q_L = \pm 1$ ($Q_U = \pm 1$) corresponds to an *up-down* (UD, upper sign) or to a *down-up* (DU, lower sign) DW configuration in the LFM (UFM) layer. $(H_x, H_y, H_z)$ are the Cartesian components of the external magnetic field. $H_D^i = \frac{D^i}{\mu_0 M_s^i \Delta}$ is the effective DMI field[18]. $H_k^i \approx M_s^i N_x$ is the magnetostatic shape anisotropy field[19], where $N_x = \frac{t_{FM}^i \log(2)}{\pi \Delta}$ is the magnetostatic factor[24]. $\alpha_i$ represents the Gilbert damping term in each FM layer, and $\gamma_0$ is the gyromagnetic ratio. Besides, the assumption that both DW widths remain constant has been made ($\Delta = \sqrt{\frac{A}{K_{eff}}}$, where $K_{eff} = K_u^L - \frac{1}{2}\mu_0(M_s^L)^2$). The term $H_{SL}^i = \frac{\hbar \theta_{SH}^i J}{2\mu_0|e|M_s^i t_{FM}^i}$ is the Slonczewskii-like term associated to the SHE, and $b_{ST}^i = \frac{\mu_B P}{|e|M_s^i} J_{FM}^i$ is the STT coefficient, with $P$ the polarization factor and $J_{FM}^i$ the density current flowing directly along the FM layers $i = \{U, L\}$. $\xi_i$ is the non-adiabatic parameter. Initially, we assume that STT is negligible ($b_{ST}^i = 0$) in the evaluated samples. We will show in Sec. 3.1.B that indeed the STT plays a marginal role on the current-driven dynamics discussed in the present study.

These 1DM Eqs. (4)-(6) can be directly expressed in the same manner as done in the supplementary information of [14]. Indeed, we verified that using the inputs considered in the supplementary material of [14], we reproduce their results (not shown). Moreover, Eqs. (4)-(6) can be used to evaluate all the cases considered here solely by proper selection of the inputs: FM coupling ($J^{ex} > 0$, with $Q_L = Q_U = \pm 1$); AF coupling ($J^{ex} < 0$, with $Q_L = \pm 1$ and $Q_U = \mp 1$), and the single FM layer case ($J^{ex} = 0$, with $Q_L = \pm 1$ and $Q_U = 0 = M_S^U = D^U = \alpha_U = \theta_{SH}^U = b_{ST}^U$). In the present work, Eqs. (4)-(6) are numerically solved using a commercial software[25].

# 3. RESULTS FOR PERFECT STRIPS

## 3.1.A Current-driven DW motion in the absence of longitudinal fields

We firstly describe the current-driven DW motion along perfect and straight systems. Representative snapshots of the local magnetization before and just at the end of a 2-ns long current pulse with amplitude $J = J_{HM} = +2.5\ \text{TA/m}^2$, are depicted in Fig. 2. In what follows, the units of the current density are given in $\text{TA/m}^2$ indicating $10^{12}\ \text{A/m}^2$. Fig. 2(b) shows the results for a single DW in the single-FM-layer stack. Fig. 2(a) and (c) correspond to a pair of DWs, one in each FM layer, equally located, in the case of the coupled multilayer system: Fig. 2(a) for FM coupling ($J^{ex} > 0$), and Fig. 2(c) for AF coupling ($J^{ex} < 0$). Except the contrary is indicated the magnitude of interlayer exchange coupling is fixed to $|J^{ex}| = 0.5 \, \text{mJ/m}^2$. Initial chiral Néel[4] configurations are stabilized in all cases, both in the single-FM-layer and in the coupled DWs of the FM coupling and AF coupling cases. In the latter case, the strong DMI at the HM/LFM interface along with the interlayer exchange interaction between the FM layers are sufficient for promoting that chiral magnetization textures. The FM coupling ($J^{ex} > 0$) makes domains in both FM layers to adopt equal orientation, leading to twin *up-down* (UD) DW transitions in both layers (Fig. 2(a)). Conversely, the AF coupling ($J^{ex} < 0$) promotes configurations where the upper and lower domains point in opposite directions. This fact results in the formation of paired DWs combining both types of DWs, one UD in one FM layer and one DU in the other FM layer (Fig. 2(c)).

The dynamical behavior of DWs in the three presented cases shows noteworthy differences. The description of such a behavior requires the definitions of the internal DW ($\Phi$) and tilting ($\chi$) angles depicted in the inset of Fig.2. Fig. 2(a) presents the results obtained for the FM coupling between the FM layers ($J^{ex} > 0$). In this case, the SOT due to the SHE acts exclusively on the magnetization of the LFM layer, pushing forward the DW in this layer. The interlayer exchange FM coupling between both FM layers results in the simultaneous displacement of the DW in the UFM layer, which is dragged by the movement of its counterpart in the LFM layer. The behavior is rather similar to the single-FM case shown in Fig. 2(b), since the inner magnetization of both DWs rotates from the initial Néel configuration similarly to the single-FM case. Note also that in these two cases the DW plane is significantly tilted due to the current. The tilting increases with $J$ and it is reduced in the FM coupling case with respect to the single FM layer case.

The case in Fig. 2(c) corresponds to the interlayer AF coupling ($J^{ex} < 0$). An antiparallel alignment of the magnetizations in the LFM and the UFM layer occurs, both within the domains and inside the DWs. Now again the movement of the DW in the LFM drags the DW in the UFM due to the AF coupling, but the highest displacements are reached. It can be checked in Fig. 2(c) that the antiparallel alignment of the magnetizations within such paired DWs almost holds during the whole dynamics, then keeping the direction of the magnetizations along the longitudinal one, that is, the parallel/antiparallel alignment of the current flow and the magnetization within the DWs, depending on the type of the DW. Interestingly, no DW tilting is observed for this AF coupling case.

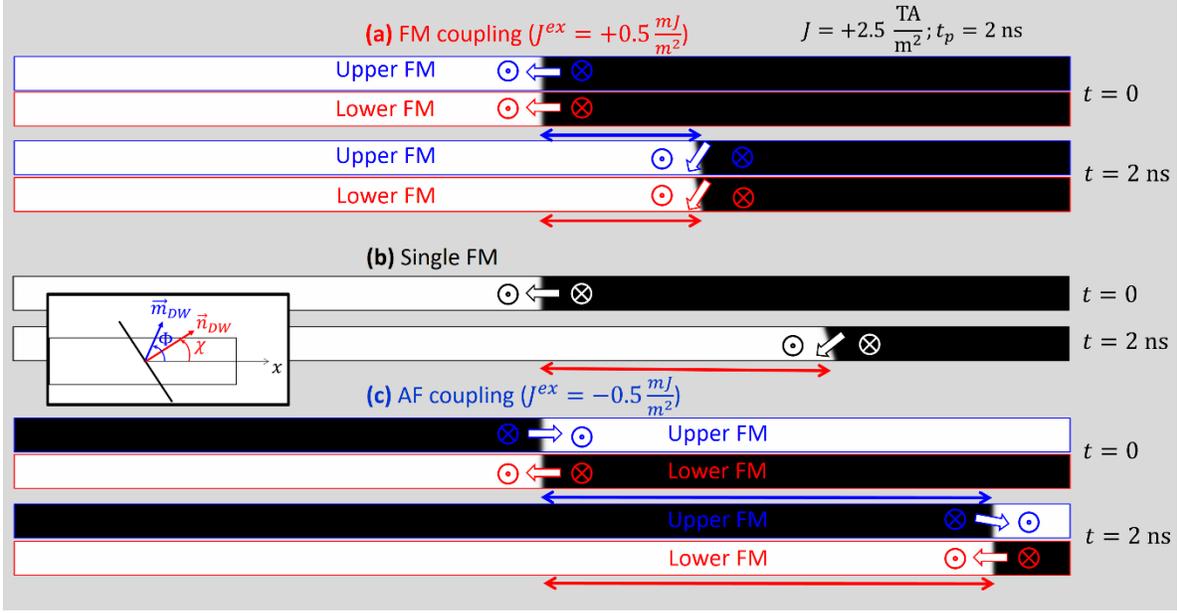

**Figure 2.** Micromagnetic snapshots of the initial ($t = 0$) and final ( at $t = 2$ ns) states of the DWs in FM layers in the following cases: (a) FM coupling ($J^{ex} > 0$), (b) single-FM-layer, and (c) AF coupling ($J^{ex} < 0$). UFM and LFM layers are simultaneously shown in cases (a) and (c). The amplitude of the current pulse along the HM layer is $J = +2.5$ TA/m$^2$. No spin current is acting on the UFM layers. Other material parameters are given in the text and in the caption of Fig. 1. Thin red arrows show the DW displacements. Black thick arrows represent the orientation of the magnetization within the DW in the single-FM-layer case, while blue and red thick arrows represent the orientation of the magnetization within the DWs in the UFM and LFM layers, respectively, for the coupled systems. Perfect samples and zero temperature are considered here. The inset depicts the definition of the DW angle $\Phi$ and the tilting angle $\chi$ as the angles formed respectively by the magnetization $\vec{m}_{DW}$ within the DW, and the normal $\vec{n}_{DW}$ to the DW, with respect to the longitudinal axis.

The DW dynamics is, in all cases, determined by a set of terminal values of the DW velocity ($v$), the DW angle ($\Phi$), the tilting angle ($\chi$) and the DW width ($\Delta$). In other words, after a short transient all these observables reach a steady-state (or terminal) regime with constant values. We checked that the terminal steady-state regime is completely reached within the first 2 ns of the current application (see Sec. 3.1.C), and therefore, this time was adopted to evaluate the terminal values of the mentioned observables. $\Phi$ and $\chi$ are computed from the terminal magnetization snapshots (at $t = 2$ ns). The DW width $\Delta$ is computed according to Thiele's definition[21]. The dependence of these terminal values on the current amplitude is shown in Fig. 3 for the three considered cases: (a) FM coupling ($J^{ex} > 0$), (b) single-FM-layer, and (c) AF coupling ($J^{ex} < 0$). In the first two cases, Fig. 3(a) and (b), the velocity asymptotically increases as the DW magnetization angle ($\Phi$) approaches a Bloch configuration (*i.e.*, a rotation of $\pm 90^0$, depending on the wall type) as $J$ increases. The variation of the internal DW angles is rather similar for both the FM coupling and the single-FM-layer cases (see graphs in Fig. 3(a) and (b)). In fact, the results of the DW magnetization angle $\Phi$ obtained for the latter case almost exactly overlap those corresponding to the LFM layer if they are plotted within the same graph (not shown). Since the SHE acts exclusively on the LFM layer, and this layer and the single-FM layer have been chosen to exactly share the same set of geometrical and intrinsic parameters, it can be concluded that the driving force due to the SHE acquires a rather similar magnitude in both cases. However, the terminal velocity ($v_{st}$) is much slower for the FM coupled system than in the latter case. This result can be understood as a lower mobility of the paired DWs in the coupled system as compared with that of the DW in the single-FM-layer. The 1DM indeed provides a clue to satisfactorily explain this lower mobility. Micromagnetic ($\mu M$) simulations show that the after a short transient the DWs adopt steady-state regime ($\dot{\Phi} = 0$) for the three evaluated cases. By imposing the steady state condition ($\dot{\Phi} = 0$) in the 1DM Eqs. (4)-(6), an analytical expression is deduced for the terminal DW velocity ($v_{st}$) of the single FM layer[18]:

$$v_{st} = \frac{\pi}{2} \frac{\gamma_0 \Delta}{\alpha} \frac{\hbar \theta_{SH} J}{2e\mu_0 M_s t_{FM}} \cos \Phi_{st}, \qquad (7)$$

where $\Phi_{st}$ is the terminal DW angle ($\dot{\Phi} = 0$, *i.e.* $\Phi \rightarrow \Phi_{st}$). Eq. (7) indicates that DW terminal velocity for a single-FM-layer monotonously increases with $J$, but with a decreasing slope as the absolute current is increased[9,18,19]. This fact can be explained by the relative orientation of the magnetization within the DWs ($\vec{m}_{DW}^i$ or $\Phi_i$ with $i$: $L, U$) and the direction of the electric current flow ($\vec{J}(t) = J(t)\vec{u}_x$). In fact, the closer the direction of the magnetization within the DWs is to the direction of the current flow, the more efficient is the SHE pushing the DWs[3,18,26]. However, the spin orbit torque (SOT) due to the SHE itself promotes the progressive misalignment of both magnetization and current: as $J$ increases, the angle $\Phi_{st}$ asymptotically tends to $90^0$ (see "DW angle *vs J*" graph in Fig. 3(b)), leading to the abovementioned decrease of the slope of the DW speed dependence on current amplitude[18,19].

Similarly, an analytical expression can be inferred from Eqs. (4)-(6) for the steady state (terminal) DW velocity in FM ($J^{ex} > 0$) and AF ($J^{ex} > 0$) coupling cases. In the cases being studied here, where $\alpha_L = \alpha_U = \alpha$, no external field is applied ($H_z = 0$), no SHE is acting on the UFM ($H_{SL}^U = 0$), no DMI on the UFM layer ($D^U = 0$) and by considering stationary conditions (*i.e.*, when $\dot{\Phi}_L = 0$ and $\dot{\Phi}_U = 0$, and $v_{st} \equiv (\dot{q})^{st}$), Eq. (4) leads to

$$v_{st} = \frac{\pi}{2}\frac{\gamma_0\Delta}{\alpha}\frac{\hbar\theta_{SH}^L J}{2e\mu_0(M_s^L + M_s^U)t_{FM}^L}\cos\Phi_L^{st},\qquad(8)$$

which, except for the term ($M_s^L + M_s^U$), is equivalent to Eq. (7) for a single FM layer. Therefore, the DW velocity is proportional to the driving current and the cosine of the stationary DW angle $\Phi_L^{st}$, and inversely proportional to a weighed-up sum of the saturation magnetizations of both FM layers, *i.e.*, the sum of $M_s^U$ and $M_s^L$. This explains the results for the FM coupling ($J^{ex} > 0$) and single FM layer cases shown in Fig. 3(a) and (b), where the variation of the DW angles are similar for both layers in the FM coupling case, and also very similar to the ones achieved for a single FM layer. However, the DW velocity is significantly larger for the single FM case as only $M_s^L$ appears in the denominator of Eq. (7), and not $M_s^L + M_s^U$ as in Eq. (8). Another difference between cases (a) and (b) can be found in the DW tilting angle ($\chi$). The FM coupling reduces the tilting angle of the paired DWs, which is rather similar for both of them, as compared with the tilting angle of the DW in the single-FM strip.

A significant contrast characterizes the CDDW dynamics in AF coupled systems ($J^{ex} < 0$). The increase of the terminal velocity of the DWs with the current amplitude is in this case rather linear in the evaluated range, and higher speeds are reached for the highest currents in the AF coupling case (see Fig. 3(c)). The key for this behavior resides in the fact that the Néel configuration of the DW in the LFM layer holds over a large range of applied currents $|J|$: the AF coupling strongly supports the antiparallel alignment of the internal DW angles (see "DW angles *vs J*" graph in Fig. 3(c)), and the SOT is not sufficiently intense to promote a significant misalignment between the current flow and the magnetization within the DWs. Actually, the use of Eq. (8) derived from the 1DM, yields a rather good approach to compute this terminal DW velocity, provided the DW angle in the LFM layer is set to $\Phi_L^{st} \approx 180^0$, as it can be seen from full $\mu M$ simulations (see "DW angles *vs J*" graph in Fig. 3(c)). However, a slight but progressive slope reduction is obtained as the current is increased, which can be ascribed to the increasing misalignment between the magnetizations within the paired DW, as the same graph reveals for high currents. Another important characteristic of this CDDW dynamics is that DW tilting completely vanishes, so that DWs hold perpendicular to the longitudinal direction ($x$-axis).

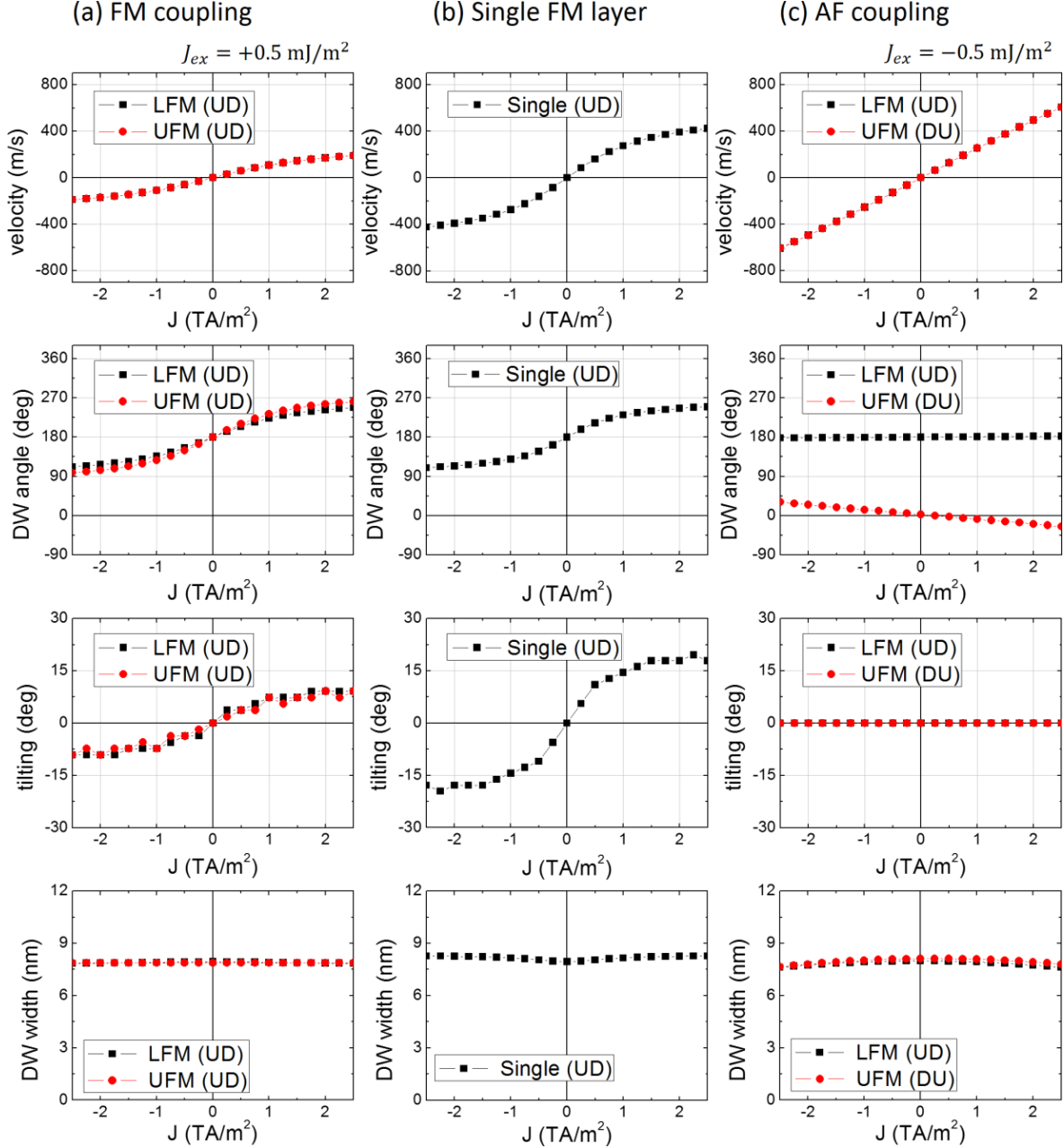

**Figure 3.** $\mu M$ results of the current driven DW motion for the three evaluated cases: (a) FM coupling ($J^{ex} > 0$), (b) single-FM layer, and (c) AF coupling ($J^{ex} < 0$). The dependence on the applied current $J$ of the terminal values of the DW velocity, the DW angle, tilting angle and the DW width, which are shown from top to bottom graphs. These values were computed at $t = 2$ ns. The parameters are those given in the text. The term UD (DU) within the legends refer to the DW magnetization transition in the LFM from *up*-to-*down* (from *down*-to-*up*) along the positive direction of the longitudinal axis ($x$-axis). Perfect samples and zero temperature conditions are considered here.

We have shown that the micromagnetic ($\mu M$) results of the DW velocity $vs\ J$ can be qualitatively described by the analytical Eqs. (7) and (8) for the single FM layer and two coupled FM layers derived from the 1DM respectively. However, it remains to check if the 1DM is also in quantitative agreement with $\mu M$ results. To evaluate it, we have numerically solved the 1DM Eqs. (4)-(6) considering the same inputs parameters as for the $\mu M$ study for the three evaluated cases: FM coupling, single-FM-layer and AF coupling. First of all, it has to be noted that the micromagnetically computed DW width ($\Delta$) dependence on $J$ shown in bottom graphs of Fig. 3 indicates that $\Delta$ almost remains independent on $J$. The $\mu M$ value agrees with the analytical prediction of DW width, which can be estimated from $\Delta = \sqrt{\frac{A}{K_{eff}}}$, where $K_{eff} = K_u^L - \frac{1}{2}\mu_0(M_s^L)^2$, resulting in a value of $\Delta \approx 7.3$ nm. The $\mu M$ results are compared to the 1DM predictions in Fig. 4.

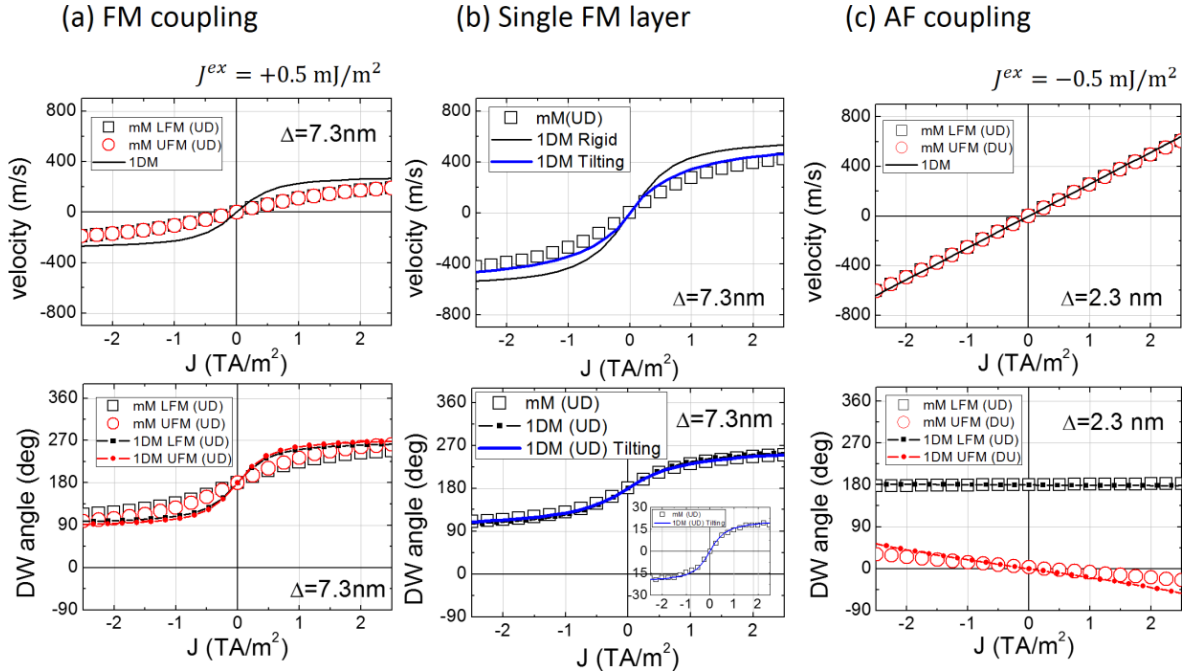

**Figure 4.** $\mu M$ results and 1DM predictions of the current driven DW motion for the three evaluated cases: (a) FM coupling ($J^{ex} > 0$), (b) single-FM-layer, and (c) AF coupling ($J^{ex} < 0$). The dependence on the applied current $J$ of the terminal values of the DW velocity and the DW angle is shown from top to bottom graphs. The parameters are those given in the text. The DW width for the AF coupling ($J^{ex} < 0$) was needed to be rescaled to $\Delta \approx 2.3$ nm in the 1DM in order achieve quantitative agreement with $\mu M$ results. For the two other cases, FM coupling ($J^{ex} > 0$) and single-FM-layer, the input value of the DW width was $\Delta \approx 7.3$ nm, as predicted by As it shown in Fig. 4, the 1DM predictions are in good qualitative agreement with the $\mu M$ results, both for the DW velocity and the DW angles. The discrepancies between the $\mu M$ and the 1DM results in the single FM layer (Fig. 4(b)) can be attributed to the approximated description provided by the 1DM, which neglects, among other aspects (such as the approximated description of the shape anisotropy field, for instance), the DW tilting observed in the $\mu M$ results. Indeed, we notice that taking into account the DW tilting in the 1DM (see Ref. [10],[18] for details) results in good quantitative agreement with the $\mu M$ data for the single-FM-layer (see blue curves in the graphs of Fig. 4(b)). Regarding the FM coupling case (Fig. 4(a)), the quantitative disagreement between 1DM and $\mu M$ results should be additionally ascribed to the magnetostatic interaction between the two FM layers, which is not taken into account in the 1DM Eqs.(4)-(6), and to the approximated description of the shape anisotropy fields ($H_k^i \approx M_s^i N_x = \frac{t_{FM}^i \log(2)}{\pi \Delta} M_s^i$)[24] considered by the 1DM. Note that this 1DM description does not take into account the width $w$ of the FM strips. On the other hand, the agreement between the $\mu M$ and the 1DM results for the AF coupling case (Fig. 4(c)) looks remarkable also from a quantitative point of view. However, this fit required to re-scale the DW width in the 1DM, which contrary to the other two cases, was set to $\Delta \approx 2.3$ nm. Note that this value is not justified by the analytical prediction ($\Delta \approx 7.3$ nm) nor by the $\mu M$ results shown in bottom graphs of Fig. 3(c). It has to be also noticed that by imposing $\Delta \approx 7.3$ nm as the input for AF coupling case, the DW velocity predicted by the 1DM overestimates the $\mu M$ results of "DW velocity $vs$ $J$" by a factor of $\sim 3$ (not shown), whereas the dependence of "DW angle $vs$ $J$" is hardly affected. For these reasons, we will continue analyzing in the following sections the current-driven DW dynamics along multilayers with two FM layers adopting a full micromagnetic description, which naturally accounts for the 3D dependence of the magnetization, including the magnetostatic interaction between them and the eventual DW tilting.

### 3.1.B The influence of the Spin Transfer Torques on the current-driven DW motion

In previous discussion we have assumed that the most of the current flows along the HM, so the only driving force on the DWs is due to the spin Hall effect (SHE), which drives DWs along the current direction for the chiral DW nature consider in the present study (left-handed chirality imposed by the DMI). However, the current could also partially flow along the FM layers, and consequently the conventional adiabatic and non-adiabatic spin transfer torques (STTs) could also contribute to the current-driven DW dynamics. In order to explore the influence of these STTs, we have evaluated the DW dynamics along the same systems studied before (Single-FM-layer stack, HM/FM/Oxide, and the multilayers with two FM layers HM/LFM/Spacer/UFM, with AF coupling) by considering that the FM layers are also submitted to the same current density as the HM, $J_{FM}^i = J_{HM}$ for $i$: $L, U$. The spin polarization factor of the STT is $P = P^i = 0.5$ for both FM layers. The geometries and materials parameters considered in the previous section have been also adopted for this analysis. The results for the terminal DW velocities as function of the current density $J = J_{FM}^i = J_{HM}$ are shown in Fig. 5(a) and (b) for the single-FM-layer stack and the multilayer HM/LFM/Spacer/UFM with AF coupling ($J^{ex} = -0.5$ mJ/m$^2$) respectively. Three different values of the non-adiabatic parameter are considered: $\xi = \xi^L = \xi^U$: 0, $\alpha = 0.1$, $2\alpha = 0.2$, and the results are compared to the ones computed in the absence of STT ($P = 0$) where the only driving mechanism is due to the SHE.

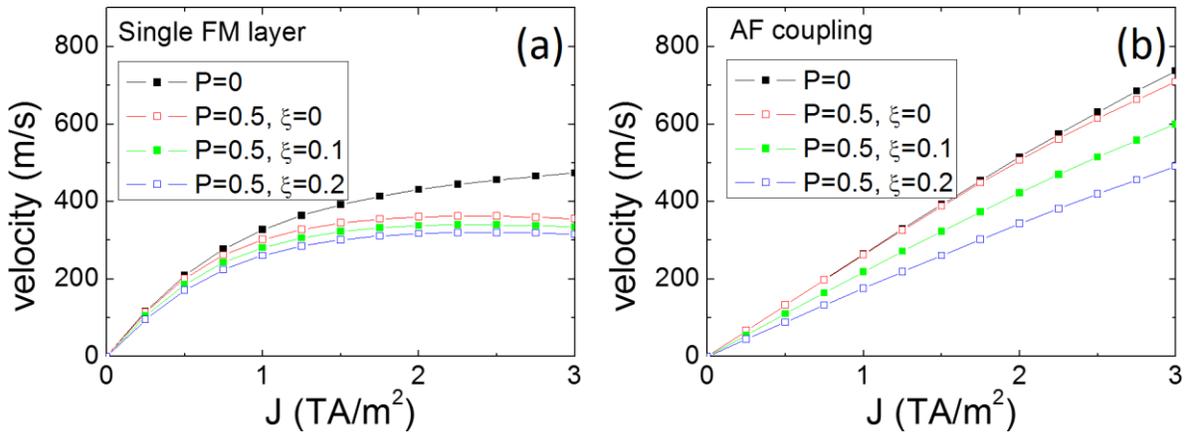

**Figure 5.** $\mu M$ results showing the dependence on the applied current $J = J_{FM}^i = J_{HM}$ of the terminal DW velocity the current driven DW motion for (a) single-FM layer stack and (b) AF coupling ($J^{ex} < 0$) in the presence of STTs. The spin polarization factor of the STTs is $P = 0.5$ for both FM layers, and different values of the non-adiabatic parameter are evaluated: $\xi = \xi^L = \xi^U$: 0, $\alpha$, $2\alpha$. The duration of the current pulse is $t_p =$

2 ns, and the velocity values correspond to the terminal state at $t = t_p = 2$ ns. Perfect samples and zero temperature conditions are considered.

The STT pushes the DW along the electron flow (against the current). As commented, for the single-FM-layer stack (Fig. 5(a)), the DW velocity due to the SHE (which drives the DW along the current direction) increases monotonously up to asymptotic saturation with $J$ in the absence of STTs ($P = 0$, black dots in Fig. 5(a)). When the STT is taken into account, the DW velocity decreases for a given current. This velocity reduction is larger as the non-adiabatic parameters increases from $\xi = 0$ (open red symbols in Fig. 5(a)) to $\xi = 2\alpha = 0.2$ (open blue symbols in Fig. 5(b)). Fig. 5(b) also shows that the DW velocity reaches a maximum for a given current, and for large currents the DW velocity starts to decrease again. These results indicate that the STTs act against the SHE, reducing the magnitude of the DW velocity, which is along the current direction.

For the HM/LFM/Spacer/UFM with AF coupling, the perfect adiabatic STT ($P = 0.5, \xi = 0$, red open symbols in Fig. 5(b)) does not significantly modify the DW velocity. Under non-adiabatic conditions ($P = 0.5, \xi > 0$, filled green and blue open symbols in Fig. 5(b)), the DW velocity decreases with respect to the zero STT case ($P = 0$). We confirmed that the DWs in the LFM and in the UFM layers move coupled even in the presence of STTs as due to the strong interlayer exchange coupling ($|J^{ex}| = 0.5$ mJ/m$^2$). Contrary to the single-FM-layer stack, the DW velocity increases monotonously with $J$, and the slope of this increasing is reduced as the non-adiabatic parameter $\xi$ increases. This indicates that the main driving force in these AF coupled multilayers is still the SHE due to the current along the HM. In the rest of the discussion we will neglect STTs for different reasons. Several experimental works[27],[28] have shown that the STT is indeed negligible in these systems. Moreover, the experiments[2] showing DW motion along the electron flow in high PMA systems are consistent with the perfect adiabatic conditions ($\xi = 0$), and this case has shown to play a marginal role for multilayers with AF coupling (Fig. 5(b)). On the other hand, we have also considered that the same current flowing through the HM is also flowing through the ultrathin FM layers. This is surely an exaggeration, as the electrical resistivity of the FM should be larger than the one of the FM layers[29],[30] ([29]: E. Martinez *et al.* Scientific Reports 5, 10156 (2015); [30]: O. Alejos *et al.* Appl. Phys. Lett. 110, 072407 (2017)). In a more realistic case, the density current along the FM layer must be smaller than the one along the thicker and low-resistivity HM, and consequently the STT should play a marginal role. For these reasons, we will not take into account the STTs in the rest of the manuscript.

### 3.1.C Inertia effect on the current-driven DW motion

In Sec. 3.1A and 3.1B, we have plotted the terminal DW velocity reached by the DWs after application of constant density current. Such values where obtained at $t_p = 2$ ns, which was found sufficient to achieve the steady-state terminal DW velocity. It is also interesting to evaluate the DW dynamics once the current pulse is turned off. The current-driven DW dynamics due to their own inertia has been studied in systems with in-plane magnetization by Thomas *et al.*[31] and Chauleau *et al.*[32], where the DW motion when the current pulse was turned off was essentially ascribed to the gyrotropic dynamics of the vortex DW configurations. Vogel *et al.*[33] shown that the DW motion induced by nanosecond current pulses in Pt/Co/AlOx multilayers with perpendicular magnetic anisotropy exhibits negligible inertia. More recent studies by Torrejon *et al.*[19] have shown that inertia effects result in a DW motion even when the current is switched off in high PMA systems with low damping. Our aim here is just to evaluate the inertia in HM/LFM/Spacer/UFM stacks, with FM ($J^{ex} > 0$) and AF ($J^{ex} < 0$) coupling, and to compare this "after-effect" to the single-FM-layer stack. To do it, we applied the current pulse at $t = 0$, and monitor the temporal evolution of the DW position and the DW velocity along perfect samples (without disorder) and at zero temperature. The results are shown in Fig. 6(a) and (b). For a single-FM-layer (open circles in Fig. 6(a) and (b)), the DW takes some time to reach its terminal velocity from $t = 0$. It also takes some time to reduce its velocity to zero once the current pulse is switched off at $t = t_p = 2$ ns. As expected, these acceleration and deceleration times increase for the FM coupling case (black squares in Fig. 6(a) and (b)) with respect the single FM layer stack. This is due to the larger effective DW mass of the FM coupled system as compared to the single-FM-layer stack[19]. Interestingly, the acceleration and deceleration times are significantly short for the system with AF coupling (blue triangles in Fig. 6(a) and (b)), which constitutes an additional advantage of these systems for some applications: DWs in these AF coupled stacks can be accelerated and decelerated faster as their single-FM-layer and FM coupled counterparts.

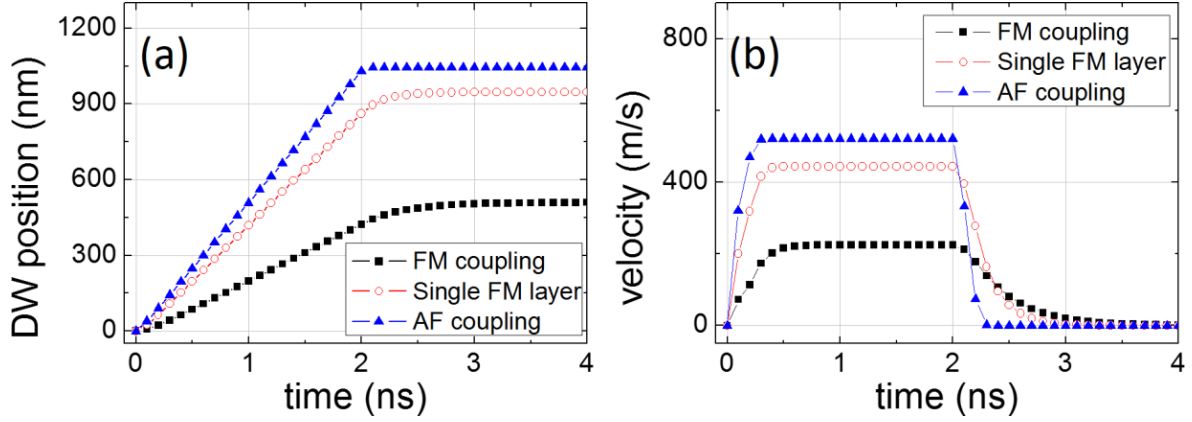

**Figure 6.** $\mu M$ results of the temporal evolution of the DW position (a) and the DW velocity (b) under a current pulse of $J = 2 \text{ TA/m}^2$ and $t_p = 2$ ns. for the three evaluated cases: FM coupling ($J^{ex} > 0$), single-FM-layer, and AF coupling ($J^{ex} < 0$). The same parameters as in Fig. 2 and 3 are considered. Depicted results correspond to perfect samples at zero temperature.

### 3.2. Current-driven DW motion under longitudinal fields

Other revealing study of the consequences of the AF coupling between the two FM layers is the dependence of DW motion on the application of an in-plane longitudinal field ($B_x$) for a given injected current. This $\mu M$ study has been then performed by taking a 2-ns long current pulse of a fixed amplitude of $J = 2.5 \text{ TA/m}^2$, with either positive ($J > 0$) or negative ($J < 0$) polarity. Again steady-state terminal values of DW observables are presented here. Within this context, positive (negative) fields mean applied fields directed along the positive (negative) $x$-axis. The three mentioned cases: FM coupling ($J^{ex} > 0$), single-FM-layer, and AF coupling ($J^{ex} < 0$) have been also evaluated. The results are shown in Fig. 7.

As it has been previously mentioned, the FM coupling between both FM layers ($J^{ex} > 0$) leads to the formation of twin DWs in both FM layers, that is, with their magnetizations perfectly aligned within the DW transition. This is a crucial point, since both magnetizations are similarly affected by the application of the longitudinal field $B_x$, either by reinforcing the alignment of the magnetization ($\vec{m}_{DW}$) and the current flow ($\vec{J}(t) = J(t)\vec{u}_x$), or by promoting their misalignment. This can be checked in the graphs of Fig. 7(a). In these graphs and the ones in Fig. 7(c), the terms UD and DU within the legends are used to refer to the magnetization transition associated to the DW in the LFM, which can go respectively from an *up*-domain to a *down*-domain (UD) and from a *down*-domain to an *up*-domain (DU), along the positive direction of the longitudinal axis ($x$-axis). According to the previous discussion, the velocity of a DU (UD) DW increases (decreases) for positive fields ($B_x > 0$) and positive currents ($J > 0$). Conversely, the velocity of a UD (DU) DW increases (decreases) for $B_x < 0$ and $J < 0$. The other cases combining different signs of the field and the current flow can be straightforwardly derived. The cases when the application of the field leads to an absolute decrease of the DW speed reach a point where the DWs freeze and no displacement occurs. Note that the FM coupling reduces the longitudinal field at which zero DW velocity is achieved with respect to the single FM layer case (compare top graphs in Fig. 7(a) and (b)). This is a clear sign of the magnetostatic interaction between the internal magnetic moments inside the DWs in the LFM and the UFM, which promotes their antiparallel alignment against the FM coupling. Further increase of the applied field magnitude promotes the inversion of the chirality of the DWs and the subsequent inversion of the direction of DW displacement. This behavior is qualitatively similar to that of a DW in a single-FM-layer (Fig. 7(b)).

Differently from this behavior, an absolute decrease of the DW velocity is obtained under the application of the longitudinal field for the AF-coupled system ($J^{ex} < 0$). As it has been shown, in the absence of driving force ($J = 0$), the magnetizations within the coupled DWs of the LFM and the UFM layers tend to be aligned antiparallel along the $x$-axis ($\Phi^L \approx 180^0$ and $\Phi^U \approx 0^0$). The longitudinal field promotes the progressive misalignment with respect to $x$-axis, independently of its sign. Therefore, due to the reduced SOT efficiency for such a misalignment, the velocity decreases as $|B_x|$ increases. In general, it can be observed that the DW tilting is not null in the presence of in-plane fields (see graphs in Fig. 7(a),(b) and (c)). Additionally, the DW width does not remain constant under $B_x$ (see bottom graphs in Fig. 7).

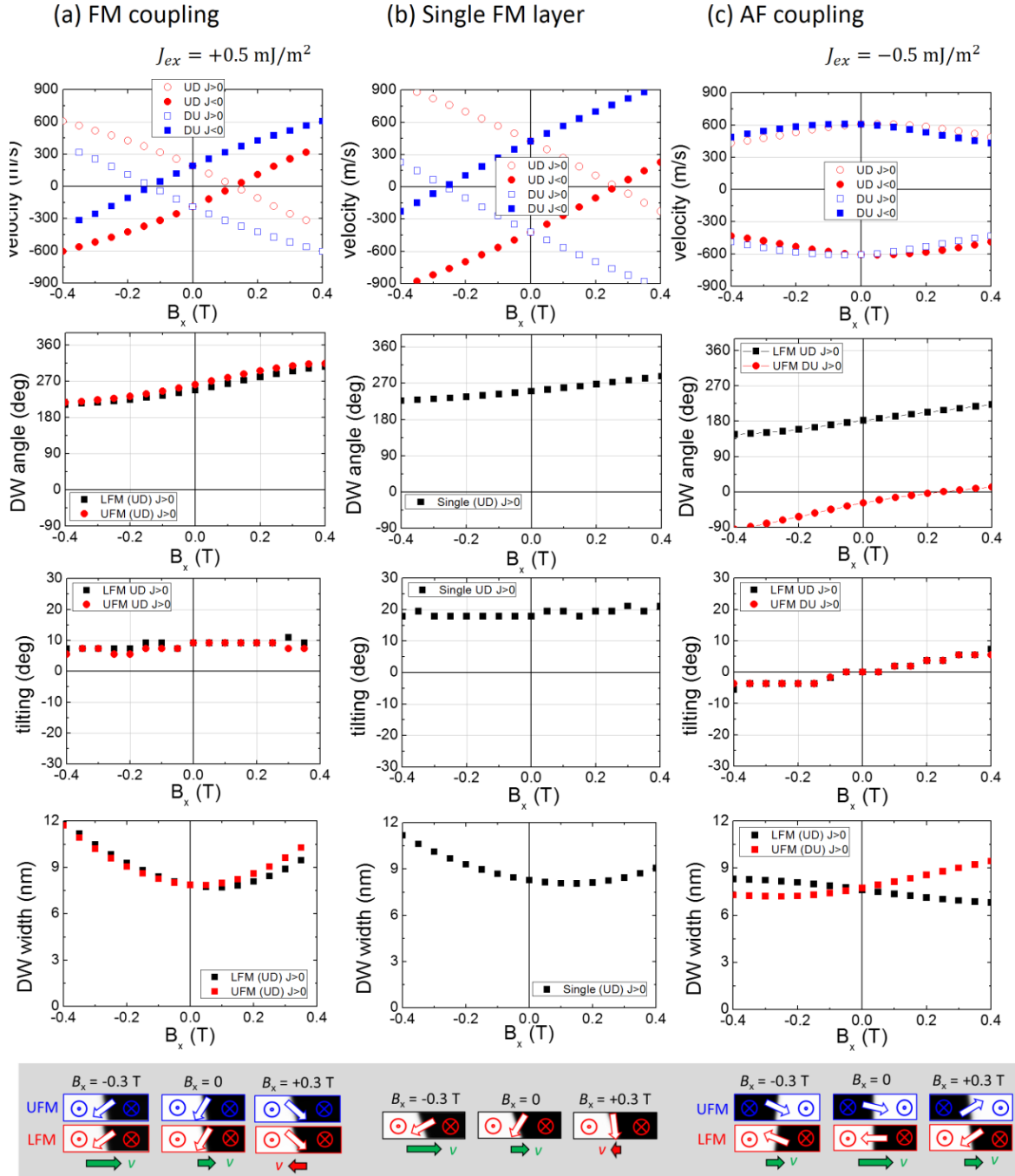

**Figure 7.** $\mu M$ results as function of the in-plane longitudinal field ($B_x$) for the three evaluated cases: (a) FM coupling ($J^{ex} > 0$), (b) single-FM layer, and (b) AF coupling ($J^{ex} < 0$). The dependence on the applied field of the terminal DW velocity, the DW angle, tilting angle and the DW width is shown from top to bottom graphs. The parameters are those given in the text. The amplitude and the duration of the current pulse are $J = 2.5 \text{ TA/m}^2$ and $t_p = 2$ ns respectively, and the presented results were computed at $t = 2$ ns when the terminal regime was already reached. The term UD (DU) within the legends refer to the DW magnetization transition in the LFM from *up*-to-*down* (*down*-to-*up*) along the positive direction of the longitudinal axis ($x$-axis). Representative snapshots are shown in bottom graphs for the three evaluated cases. Perfect samples and zero temperature conditions are considered.

We have also evaluated the 1DM predictions for the current-driven DW motion in the presence of longitudinal fields. The 1DM results are collected and compared to the $\mu M$ results in Fig. 8. A good qualitative agreement is achieved for the three cases. The quantitative discrepancies are due to the same limitations discussed above for the pure current-driven case: the 1DM does not take into account the DW tilting angle nor the magnetostatic interaction between the two FM layers. Moreover, it assumes that the DW width is fixed, which is not the case of the full $\mu M$ results shown in the bottom graphs of Fig. 7. Nevertheless, the 1DM gives a good description of the $\mu M$ results provided that the DW width ($\Delta = 2.3$ nm) is properly selected for the AF coupling case. The agreement is also good for the FM coupling and single-FM-layer cases adopting a constant DW width as deduced from the analytical formula $\Delta = \sqrt{A/K_{eff}} = 7.3$ nm.

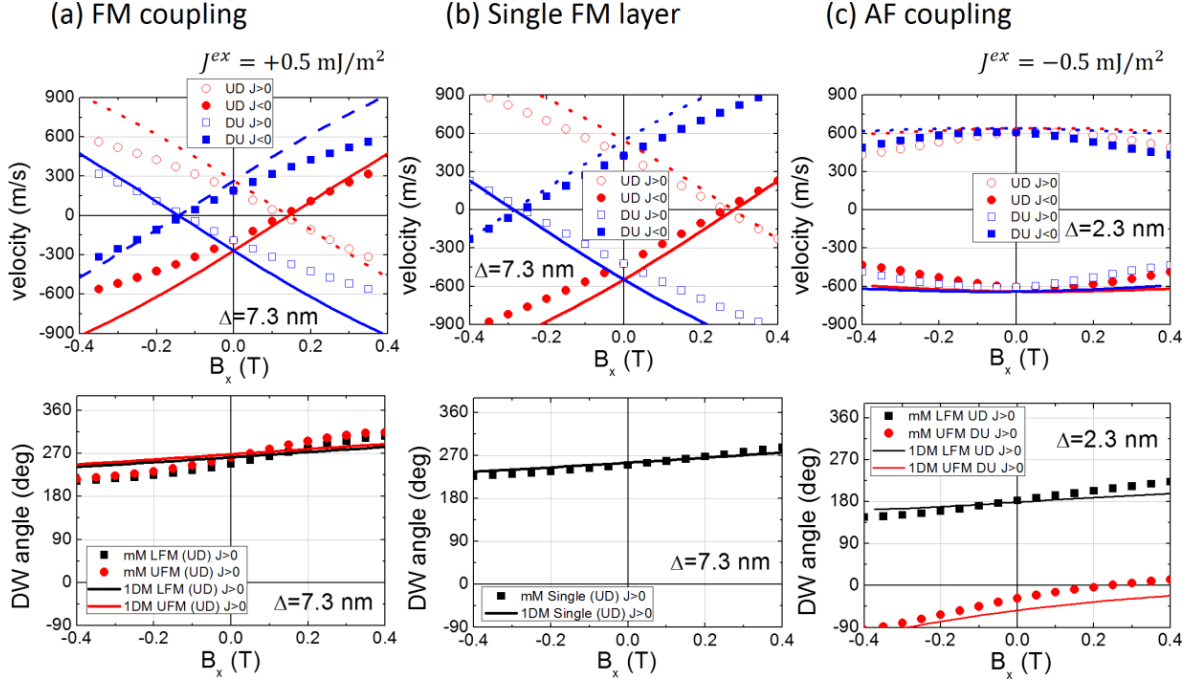

**Figure 8.** $\mu M$ results (dots) and 1DM (both solid and dashed lines) predictions of the current driven DW motion under longitudinal fields ($B_x$) for the three evaluated cases: (a) FM coupling ($J^{ex} > 0$), (b) single-FM layer, and (b) AF coupling ($J^{ex} < 0$). The dependence on $B_x$ of the DW velocity and the DW angle is shown from top to bottom graphs. The parameters are those given in the text. The amplitude and the duration of the current pulse are $J = 2.5$ TA/m$^2$ and $t_p = 2$ ns respectively, and the presented results were computed at $t = 2$ ns. The DW width for the AF coupling ($J^{ex} < 0$) was needed to be rescaled to $\Delta \approx 2.3$ nm in the 1DM in order achieve quantitative agreement with $\mu M$ results. For the two other cases, FM coupling ($J^{ex} > 0$) and single-FM layer, the input value of the DW width was $\Delta \approx 7.3$ nm, as predicted by the analytical formula $\Delta = \sqrt{A/K_{eff}}$. Results correspond to perfect samples and zero temperature.

### 3.3. Current-driven DW motion as a function of the interlayer exchange coupling

Before discussing the case of realistic samples with imperfections, it is interesting to examine the current-driven DW dynamics for different values of the exchange coupling between the layers ($J^{ex}$). The $\mu M$ results of the DW velocities of the lower and the upper FM layers are shown in Fig. 9 for two different current density amplitudes $J$, and for two different combinations of the saturation magnetization in the LFM and the UFM layers: (a) $M_s^L = M_s^U = 600$ kA/m, and (b) $M_s^L = 600$ kA/m and $M_s^U = 800$ kA/m. Gray rectangle indicates the range of $J^{ex}$ where the DWs in the LFM and UFM move uncoupled from each other, *i.e.* DWs in the LFM and in the UFM depict different velocities. For strong interlayer coupling, the DWs move coupled, but for small $|J^{ex}|$ they move uncoupled. The range of uncoupled DW motion is different for both evaluated cases, and it is wider when the FM layers have different saturation magnetization. Note also that this uncoupled range is not symmetric with respect to $J^{ex} = 0$. The fact that the threshold magnitudes of the interlayer exchange coupling needed for the coupled DW dynamics is different from the FM ($J^{ex}>0$) and AF ($J^{ex}<0$) coupling cases indicates that indeed the magnetostatic coupling between the layers plays a role in the DW dynamics. This magnetostatic interaction between the magnetization in the FM layers is complex in general. It includes different contributions: Domain-Domain ($\vec{H}_{d,D-D}^{i \to j}$), Wall-Domain ($\vec{H}_{d,W-D}^{i \to j}$) and Wall-Wall ($\vec{H}_{d,W-W}^{i \to j}$) interactions where $i,j: L, U$. (for example, $\vec{H}_{d,W-W}^{L \to U}$ represents the magnetostatic interaction generated by the internal DW magnetic moment in the Lower DW on the Upper DW). These interactions, which in general are difficult to isolate from the global magnetostatic interaction in the system, are schematically shown in Fig. 9(c) and (d) for and AF ($J^{ex}<0$) and FM ($J^{ex}>0$) coupling cases. Note that even in the symmetric case ($M_s^L = M_s^U = 600$ kA/m) there is not a complete compensation for AF coupled layers (Fig. 9(c)): although the Wall-Wall magnetostatic interaction ($\vec{H}_{d,W-W}^{i \to j}$) supports the AF coupling, the Domain-Domain ($\vec{H}_{d,D-D}^{i \to j}$) interaction not. For the FM coupling case (Fig. 9(d)), the magnetostatic interaction between the wall moments ($\vec{H}_{d,W-W}^{i \to j}$) acts against the exchange interlayer coupling, which promotes their parallel alignment. On the contrary, the parallel alignment of the magnetization in the Lower and the Upper Domains is assisted by this magnetostatic interaction ($\vec{H}_{d,D-D}^{i \to j}$).

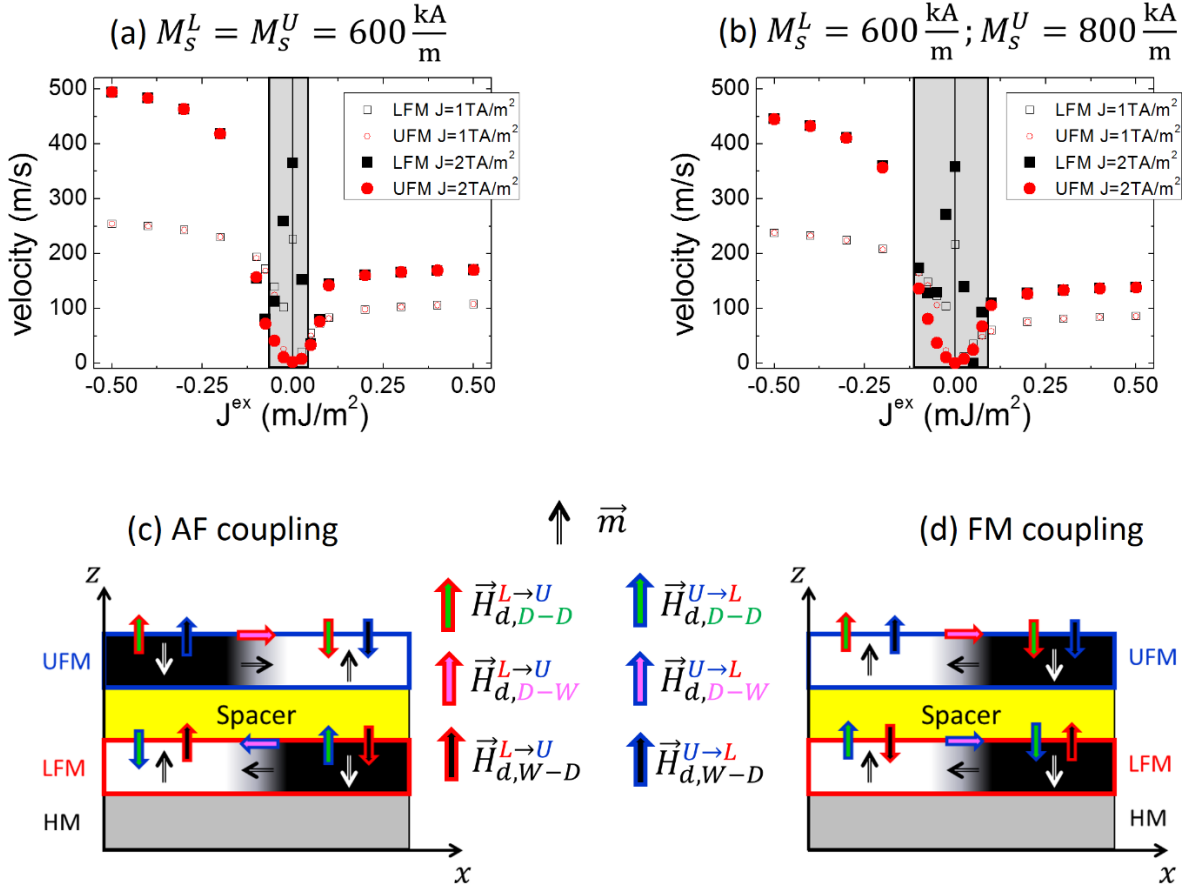

**Figure 9.** $\mu M$ results of the current driven DW motion for different values of the exchange coupling parameter $J^{ex}$ and two applied currents of amplitudes $J = 1\,\mathrm{TA/m^2}$ or $J = 2\,\mathrm{TA/m^2}$. Two different combinations of the saturation magnetization in the LFM and UFM are considered: (a) $M_S^L = M_S^U = 600\,\mathrm{kA/m}$, (b) $M_S^L = 600\,\mathrm{kA/m}$ and $M_S^U = 800\,\mathrm{kA/m}$. The gray range indicates the range of $J^{ex}$ where the DWs in the LFM and UFL move uncoupled from each other. Results were obtained at zero temperature. (c) and (d) show a schematic representation of the magnetostatic field created by magnetization in the $i$ layer on the $j$ layer for the AF and FM coupling cases respectively. Domain-Domain ($\vec{H}_{d,D-D}^{i \to j}$), Wall-Domain ($\vec{H}_{d,W-D}^{i \to j}$) and Wall-Wall ($\vec{H}_{d,W-W}^{i \to j}$) are shown.

# 4. MICROMAGNETIC RESULTS FOR REALISTIC AND ASYMMETRIC MULTILAYERS

Most of the former results were obtained for perfect samples considering two FM layers with identical thickness ($t_{FM}^L = t_{FM}^U = 0.8\,\mathrm{nm}$) and saturation magnetization ($M_S^L = M_S^U = 600\,\mathrm{kA/m}$). The thickness of the spacer was also equal to the one of the FM layers ($t_{FM}^L = t_S = t_{FM}^U = 0.8\,\mathrm{nm}$). In this section, we study the current-driven DW motion along realistic strips, *i.e.* with imperfections (details were given at the end of Section 2.1), and considering different FM layers, with different thicknesses ($t_{FM}^L$ and $t_{FM}^U$) and saturations magnetization ($M_S^L$ and $M_S^U$). The thickness of the spacer is also varied ($t_S$). Besides the FM coupling ($J^{ex} > 0$), single-FM-layer and AF coupling ($J^{ex} < 0$) cases, the $\mu M$ results collected in Fig. 10 also include the case where the two FM layers are not exchange coupled ($J^{ex} = 0$, red circles). Note that in the absence of interlayer exchange coupling, only the DW in the LFM is displaced as due to the SHE, which, as already mentioned, in the present work is only acting in the LFM layer. Therefore, red circles in Fig. 10 correspond to the DW velocity in the LFM layer, whereas black squares (FM coupling) and blue triangles (AF coupling) represent the DW velocities in both the LFM and the UFM layers, where they move coupled.

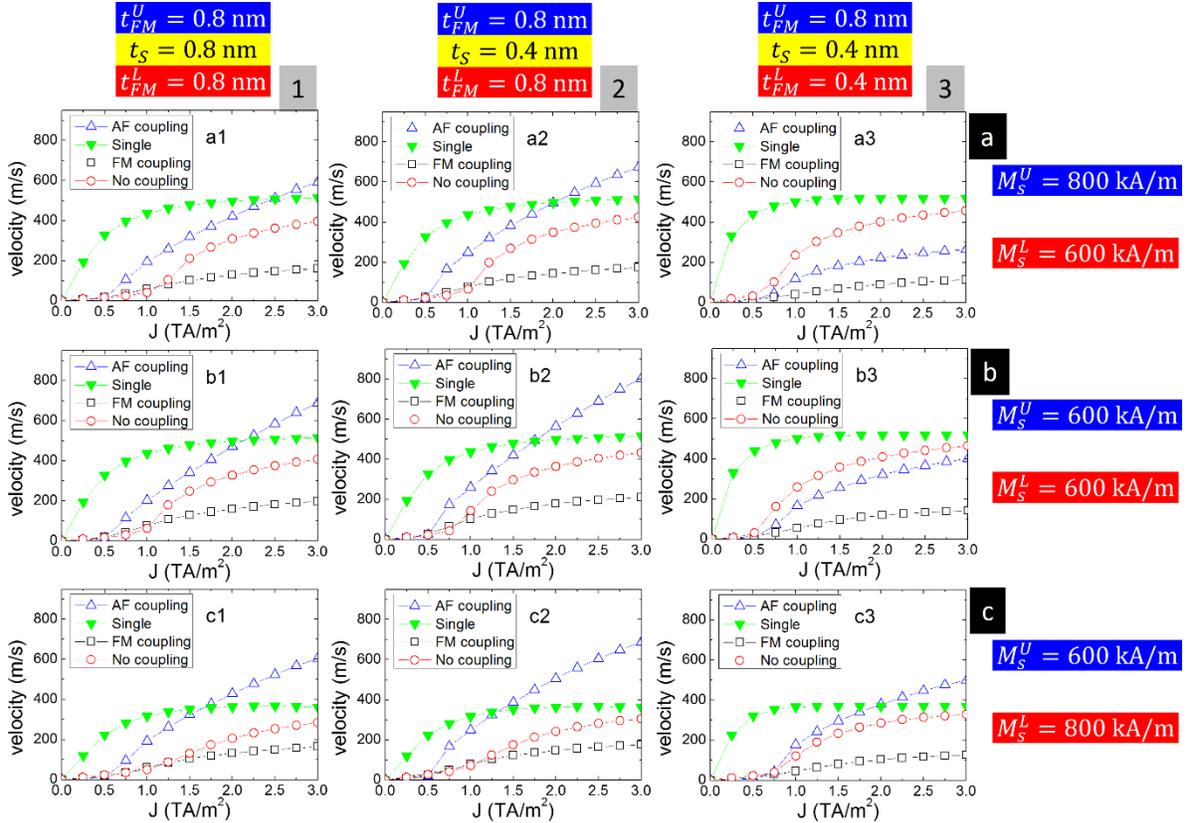

**Figure 10.** $\mu M$ results of the current driven DW motion along realistic strips for different combinations of the saturation magnetization in the FM layers (a,b,c, from top to bottom) and different thicknesses of the layers (1,2,3, from left to right). The magnitude of the exchange coupling parameter is $|J^{ex}| = 0.5$ mJ/m² for the FM ($J^{ex} > 0$) and AF ($J^{ex} < 0$) coupling cases and zero for the no coupling case ($J^{ex} = 0$). These results were obtained at zero temperature for realistic samples, with defects included as described in Sec. 2.

Several important conclusions can be extracted from the results shown in Fig. 10. First, the DW in the single-FM-layer case (green triangles) is less sensitive to the imperfections. Such imperfections introduce a propagation threshold ($J_P$) for the DW motion in the other cases (HM/LFM/Spacer/UFM), where the DW dynamics is only driven by the SHE in the HM and the interlayer exchange coupling. In other words, the SHE, acting only in the LFM, must overcome the DW pinning in both FM layers. In the low current regime (which is limited by the combination of thicknesses ($t_{FM}^L$, $t_{FM}^U$ and $t_S$) and saturation magnetization values ($M_S^L$ and $M_S^U$)), the single-FM-layer case is the one depicting higher velocities. However, in the high $J$ regime, the DW velocity saturates, as it was already explained here and elsewhere[9,18]. For higher values of $J$, (except for a3 and b3 dominated by a strong magnetostatic interaction between the internal DW moments), the largest velocity is achieved for the AF coupling case (blue triangles), and the smallest one in the FM coupling case (black squares). Note, that the DW velocity increases monotonously with $J$ in the AF coupling case. The influence of the disorder is evident for all cases, and in general, the propagation threshold is magnified in the absence of exchange coupling between the FM layers (see red circles for $J^{ex} = 0$). This is expected again as due to the magnetostatic interaction between the two dissimilar FM layers. In fact, this interaction promotes the antiparallel alignment of the internal magnetic moments, and therefore it results in an attractive force between these two DWs, which naturally explains the larger propagation threshold of the DW in the LFM in the absence of exchange coupling ($J^{ex} = 0$). For the AF coupling case ($J^{ex} < 0$) in the high current regime, the largest velocity is reached when the thickness of the spacer ($t_S$) is reduced (compare cases 1 and 2 in Fig. 10). Again, this is a consequence of the magnetostatic interaction: as $t_S$ is reduced, the dipolar interaction between the internal magnetic moments supports their antiparallel alignment resulting in a larger DW velocity. Also, the DW velocity increases as the saturation magnetization of the FM layers is equal ($M_S^L = M_S^U = 600$ kA/m, compare cases a, b and c in Fig. 10). This later fact was already qualitative explained by Eq. (8), where the DW velocity scale with $(M_S^L + M_S^U)^{-1}$.

# 5. CURRENT-DRIVEN MOTION ALONG CURVED STRIPS

Apart from the larger velocity of the DWs, another important advantage of using AF coupled layers with respect to the single-FM-layer stacks is the absence of DW tilting (see Fig. 2 and 3). In a single-FM-layer stack, adjacent DWs depict opposite tilting of their DW plane[34,35], which imposes a limit in the density of information coded between adjacent DWs. Indeed, the DW tilting can result in the annihilation of adjacent DWs, leading to mischievous effects on the coded information in a DW-based device. Contrarily, our present study indicates that when two FM layers are antiferromagnetically coupled, the DWs are driven by the current without significant tilting. For possible applications, trains of DWs must be displaced by the action of the current not only along straight paths, but also along curved paths. Accordingly, the motion of trains of DWs along both a single-FM-layer and a multilayer with two AF coupled FM layers have been separately studied.

The case of the DW displacement along FM curved strips constitutes one of the most interesting examples of application of the previous study. DW tilting limits in much cases the feasibility of these elements as racetrack memories[36], since DWs in these strips move at different velocities at the curved sections, depending on its UD or DU configuration. As an example of this curved geometries, a strip composed of three straight sections and two round-shaped sections, *i.e.*, an inverse S-shaped element was evaluated. Its geometry is depicted in Fig. 11 (a). The evaluated dimensions are given in the caption of Fig. 11.

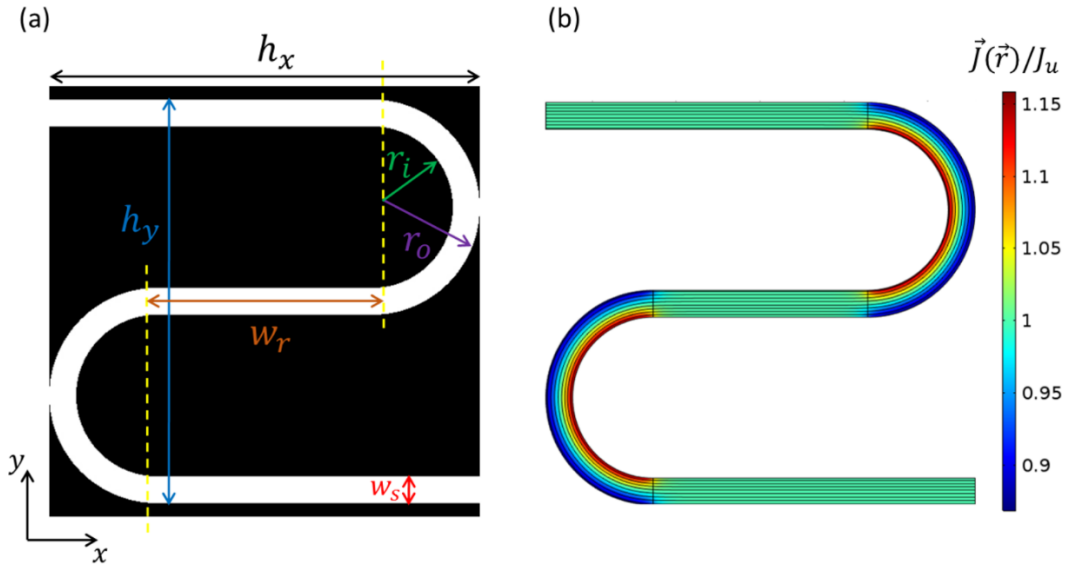

**Figure 11**. In-plane geometry to evaluate the current-driven DW dynamics along samples with curved parts. The strip contains three straight sections connected by two round-shaped sections. (a) The values of the geometrical parameters defined therein are: $w_r = 512$ nm, $r_i = 192$ nm, $r_o = 256$ nm, $h_x = 2r_o + w_r$, and $h_y = 3r_o + r_i$. (b) Spatial distribution of the normalized current density $(\vec{J}(\vec{r})/J_u)$ along the heavy metal (HM) under the lower FM layer. Color indicates the current density $(\vec{J}(\vec{r}))$ normalized to the value in the straight part $(J_u)$, where the current is uniform $J_u$.

The current density $\vec{J}(\vec{r})$ becomes non-uniform when is forced to flow along curved paths. The current distribution in the HM under the lower FM layer is shown in Fig. 11(b), which clearly indicates a radial dependence: the current density $\vec{J}(\vec{r})$ depicts an inversely linear dependence on the radius when is forced to flow over semicircular arcs. These results were computed with COMSOL[22] and taken into account to evaluate the current driven DW motion. Realistic conditions have been considered, which include defects in the form of grains (see details at the end of Section 2.1) and thermal effects at room temperature ($T = 300$ K). Two cases are considered, a single-FM-layer stack (Fig. 12) and a multilayer with AF coupling (Fig. 13). In both cases, a series of DWs is initially placed at one of their ends (see the areas surrounded by a green dashed rectangle at the upper ends of the strips), then defining a set of *up* and *down* domains in identical configuration along both strips. Currents of amplitude $J = 2$ TA/m$^2$ run and push forward the series of DWs in each strip.

Snapshots of the displacement of the DWs are depicted in Fig. 12 for a single-FM-layer. A DW annihilation event starts at a time $t = 4$ ns and takes place at the beginning of the lower curved section (see the *down* domain surrounded by the solid red square). The DWs limiting such a reduced *down* domain are moving from right to left in the preceding straight section, but they reach different velocities as they enter the curved section[36]. This fact, together with the inherent tilting of both DWs, result in the DWs making contact at their upper end, and then mutually annihilating. Ten different stochastic realizations of the thermal noise were evaluated, all of them showing similar annihilation events.

Similar DW trains are also considered in the multilayer with AF coupling. The results are shown in Fig. 13. The initial configuration of the DW train holds in both FM layers along the whole dynamics, so that this DW train successfully reaches the bottom-lower end without annihilation. Differently from the preceding case, the formation of paired DWs in the LFM and UFM layers, containing both types of DWs, UD and DU, results in equalized velocities along the curved sections for the two types of magnetization transitions in these strips. Additionally, the absence of DW tilting reduces the likelihood of a contact between adjacent DWs.

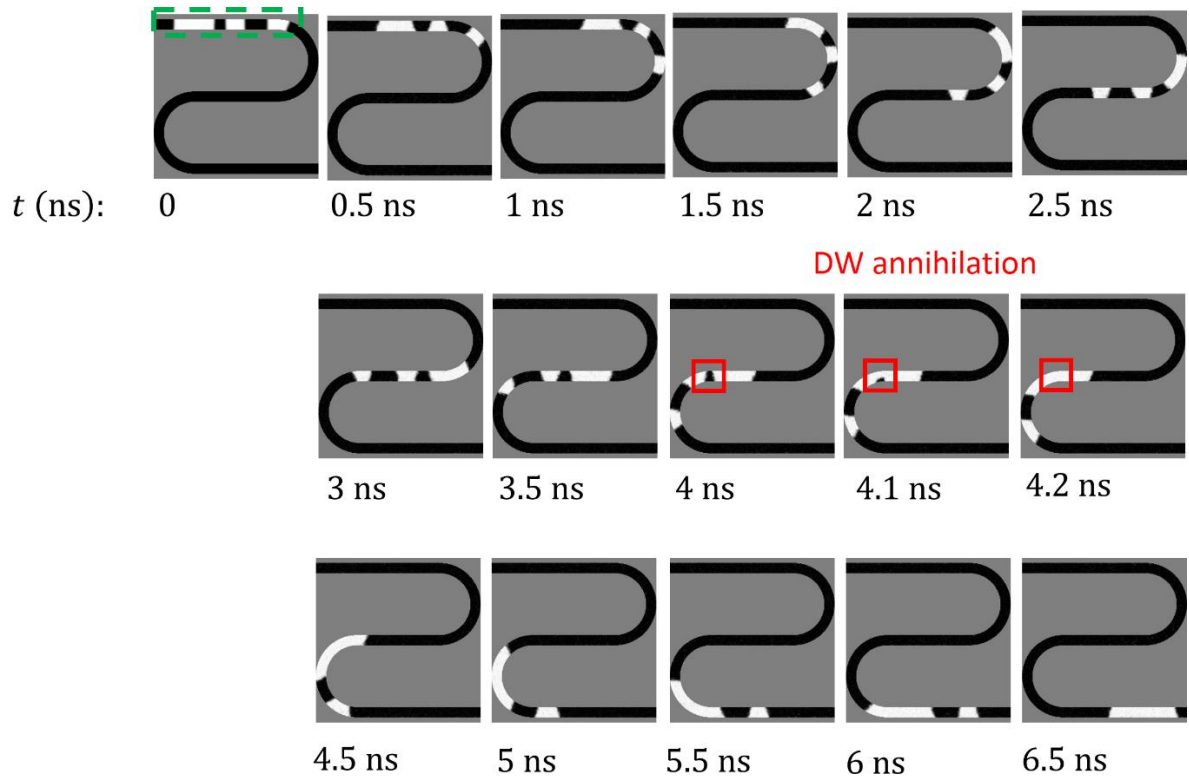

**Figure 12.** Displacement of a DW train along a curved strip in a single-FM strip. The initial state consists of a given DW train as defined in the upper straight surrounded by a dashed green rectangle. The snapshots show the displacement in time of the set of DWs under the application of a current with amplitude $J = 2\,\mathrm{TA/m^2}$ at a temperature of $T = 300$ K. Realistic samples with defects are considered here. A DW annihilation process starts at a time around $t = 4$ ns in the area surrounded by a solid red square.

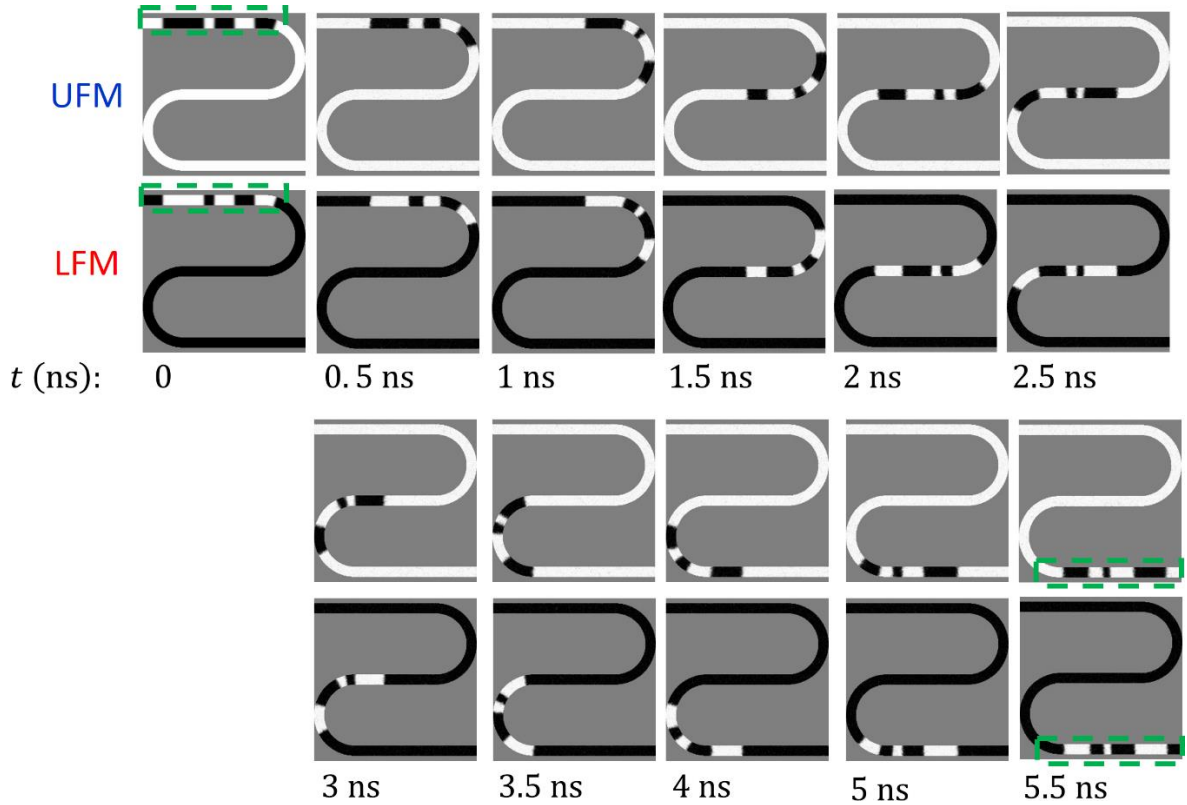

**Figure 13.** Displacement of a DW train along a curved strip corresponding to the AF-coupled multilayer. Two analogous DW trains as in Fig. 12 are considered as initial state in the upper and lower FM layers, as shown within the dashed green rectangle in the upper ends of the UFM and LFM strips. The snapshots show the displacement in time of the set of DWs under the application of a current with amplitude $J = 2\,\text{TA/m}^2$ at a temperature of $T = 300$ K. Realistic samples with defects are considered here. The train of DWs is displaced with the same velocity along the straight and curve parts of the strips, and no DW annihilation is observed.

In principle, besides of the high efficient DW dynamics, AF-coupling systems can also improve the density of packed information, coded between adjacent walls. However, a deeper observation of the images shown in Fig. 13 indicates that the second *down* domain in UFM (second *up* domain in the LFM) is contracted when arriving at the first curve (third image in Fig. 13, at $t = 1$ ns). Then, this *down* domain extends a little bit on the straight line and again contracts at the second curve. Therefore, it seems that under realistic conditions (defects and thermal fluctuations) the distance between adjacent DWs can also vary during the motion even for the AF coupling case, and consequently it is needed to evaluate the distance between adjacent DWs for realistic conditions. In order to get further insight into this behavior, we have also evaluated the dynamics of two DWs within each FM layer starting from different distances between them ($d_0$). We monitor the evolution of the distance between these DWs at five different points along the curved sample. These points are labeled with letters in Fig. 14(a), which corresponds to an initial state where two DWs initially separated by $d_0 \approx 130$ nm. The snapshots shown in Fig. 14(a) were obtained in the presence of disorder (see disorder details in Sec. 2.1) but at zero temperature ($T = 0$). It can be visually checked that the initial distance between the 2 DWs is not changing as they are driven along the track (see also Fig. 15(a)). However, in the presence of thermal noise at $T = 300$K, we notice that the distance between the DWs slightly changes: see Fig. 14(b) and (c), which correspond to two different stochastic realizations and the same grain pattern. We recorded the temporal evolution of the out-of-plane component of the magnetization at the same five points indicated in Fig. 14(a): $m_z(i,t)$ with $i: A, B, C, D, E$. The results of $m_z(i,t)$ vs $t$ corresponding to the cases depicted in Fig. 14 (a),(b) and (c) are shown in Fig. 15(a), (b) and (c). In order to provide a quantitative estimation of the temporal evolution of the distance between the DWs, we computed the difference in the switching (DW passage) times at the mentioned points: $\Delta t_S \equiv t_S^{DU} - t_S^{UD}$, where $t_S^{DU}$ and $t_S^{UD}$ correspond the times at which the left (DU) and the right (UD) DWs pass across the mentioned points. This temporal interval $\Delta t_S$ constitutes a measure of the distance between the DWs as they are driven along track. As it is shown in Fig. 15(d), $\Delta t_S$ does not vary from point to point at zero temperature (blue dots in Fig. 15(d)).

In order to provide a statistically description of this thermally activated dynamics, we evaluated three different grain patterns and three different stochastic realizations of the thermal noise at $T = 300$K. The corresponding results of $\Delta t_S$ at the mentioned points are shown by open symbols in Fig. 15(d), which indicates that the distance between the walls changes for different grains patterns and temperature realizations. However, the mean distance averaged over these grains patterns and stochastic realizations (red squares in Fig. 15(d)) is hardly dependent on the point along the track. We have performed a similar study starting from two DWs initially separated by $d_0 \approx 60$ nm, and we verified that the DWs can collapse for some of the evaluated stochastic realizations. Therefore, this imposes a limit in the density of packed information even for the AF coupling stacks. Although further studies are needed to evaluate other samples with different strip width and curvature radius, our analysis suggests that the AF coupled multilayers could be used to efficiently drive trains of highly packed DWs.

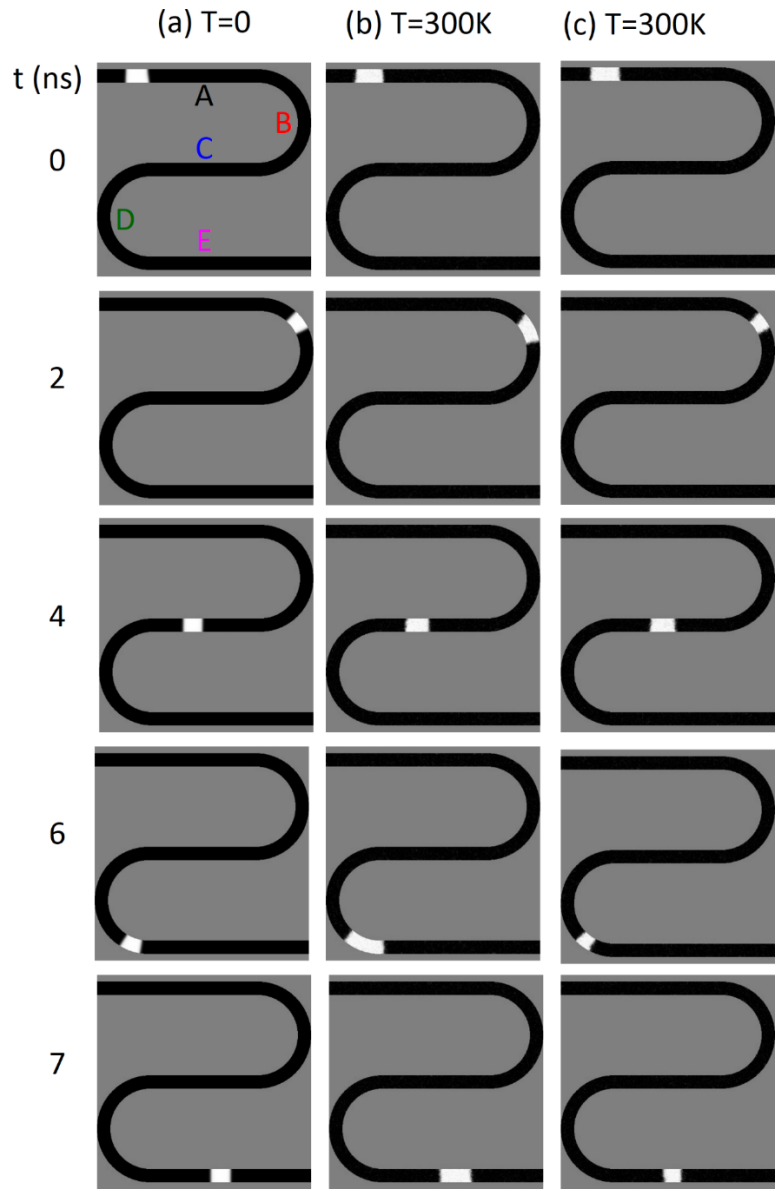

**Figure 14.** Displacement of two DWs along the AF-coupled multilayer under the application of a current with amplitude $J = 2 \, \mathrm{TA/m^2}$. The snapshots correspond to the LFM and show the temporal displacement of two DWs: (a) Sample with disorder in the form of grains at zero temperature. (b) and (c) correspond to two different stochastic realizations computed at room temperature for the same grain pattern as in (a).

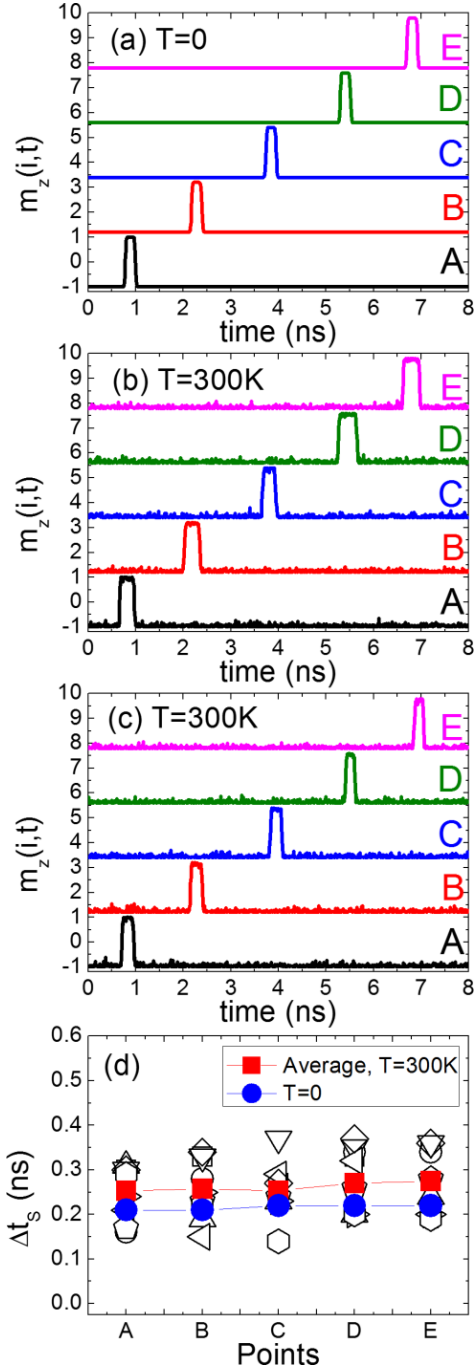

**Figure 15.** Temporal evolution of the out-of-plane magnetization ($m_z(i,t)$) at five different points ($i$:A,B,C,B,E) along the LFM layer of the AF-coupled multilayer under the application of a current with amplitude $J = 2$ TA/m$^2$ for three cases: (a) sample with disorder and at $T = 0$. (b) and (c) correspond to the same grain pattern as in (a) but for two different stochastic realizations of the thermal noise at $T = 300$K. (a), (b) and (c) correspond to the snapshots shown in Fig. 14. (d) $\Delta t_S = t_S^{DU} - t_S^{UD}$ as defined in the text at different points (A,B,C,B,E) along the LFM layer of the AF-coupled multilayer under the application of a current with amplitude $J = 2$ TA/m$^2$. These points are marked in Fig. 14. Open symbols correspond to different grain patterns and different stochastic realizations of the thermal noise. Red symbols depict the average over grains patterns and thermal realizations, and blue symbols are zero temperature results.

## 6. CONCLUSIONS

The current-driven DW motion has been studied by micromagnetic simulations in multilayers with two ferromagnetic layers separated by a spacer. These layers are coupled by the interlayer exchange coupling, which depending on its magnitude and sign, can generate ferromagnetic (FM) or antiferromagnetic (AF) coupling between them. The interfacial Dzyalozinskii-Moriya interaction is only required at the interface between the heavy metal layer and the lower ferromagnetic layer, and provides the magnetization domain wall texture with the adequate chirality. The results are compared to the ones obtained for the single-ferromagnetic-layer case and qualitatively explained in terms of analytical expressions deduced from the one-dimensional model. However, the three-dimensional space micromagnetic description allows for unraveling some details of such dynamics that are not fully accessible from a one-dimensional description, even though the latter approach may draw rather good qualitative results. For low currents in perfect samples, the driving force resulting from spin-orbit torques (spin Hall effect) is not capable to impel paired walls as efficiently as domain walls in the single-ferromagnetic-layer stack. Indeed, domain walls in the upper ferromagnetic layer are dragged by the moving walls in the lower ferromagnetic layer, because of the interlayer exchange coupling, which results in this lack of effectiveness. For higher currents, the coupled walls associated to the FM coupling present an analogous behavior to that of domain walls in the single layer stack, *i.e.*, the domain wall velocity saturates as the current is increased.

On the other hand, the AF coupling results in a high velocity of the coupled DWs, which are driven without significant tilting by the spin Hall effect from the heavy metal. The antiferromagnetic coupling promotes the antiparallel alignment of the internal DW moments in the lower and in the upper layers, both depicting a chiral Néel configuration. As consequence of that, the DW increases monotonously with current density. The velocity of the AF coupled DWs is enhanced as the saturation magnetization of the layers become similar in magnitude, and when their values decrease. Full realistic micromagnetic simulations indicate a faster coupled DW motion when the thickness of the spacer between the FM layers is reduced, and also when these layers exhibit equal saturation magnetization. While this later observation can be qualitatively described by the simple one-dimensional model, the first one is a direct consequence of the magnetostatic interaction between the internal magnetic moments of the DWs, which supports the antiparallel orientation between the internal magnetic moments in the AF coupling case. The conventional spin transfer torques does not significantly perturb the current-driven DW dynamics generated by the Slonczewski-like spin-orbit torque in AF coupled stacks, at least under perfect adiabatic conditions. It was also observed that inertia effects are significantly reduced in AF coupled stacks with respect to the single-FM-layer and FM coupling cases. The high efficiency of the current-driven DW dynamics in these AF systems is also coherent with the results obtained under in-plane longitudinal applied fields, which are also presented here.

Our micromagnetic simulations also indicate that *up-down* and *down-up* domain walls move with different velocities along a single-FM-layer stack with curved parts. Moreover, domain wall tilting constitutes another important issue that interferes with the proper working of DW-based racetrack memories. This is particularly critical in the case of single-FM-layer stacks with curved parts, since this tilting may give rise to domain wall annihilation, and consequently, imposes a limit for the high density packing of domain walls. Our micromagnetic simulations have also revealed antiferromagnetic coupling as a sound ally to avoid tilting and, consequently, to help the safe displacement of domain walls along such curved geometries. For these antiferromagnetic coupled stacks, *up-down* and *down-up* walls move with the same velocity along curved tracks at zero temperature. However, very close DWs can collapse even for AF coupling stacks under realistic conditions. The variation of the relative distance between adjacent walls is due to thermal fluctuations. Therefore, further systematic theoretical and experimental studies are needed to evaluate this limitation for strips with different widths and curvature radius.

**Acknowledgments**

This work was supported by project WALL, FP7- PEOPLE-2013-ITN 608031 from the European Commission, project MAT2014- 52477-C5-4-P from the Spanish government, and project SA282U14 and SA090U16 from the Junta de Castilla y Leon.